\documentclass[onecolumn,superscriptaddress,secnumarabic,amssymb,amsmath,nobibnotes,aps,prd,nofootinbib,altaffilletter]{revtex4}
\usepackage{amsmath,amssymb}
\usepackage{graphicx}
\usepackage{dcolumn}
\usepackage{bm,color}
\usepackage{hyperref}
\usepackage{accents}
\usepackage{amssymb,float}
\usepackage{amsmath}
\usepackage{multirow}
\usepackage{siunitx}
\usepackage{tabularx}
\usepackage{booktabs}
\usepackage{url}
\usepackage{capt-of} 
\usepackage{placeins} 
\usepackage{psfrag}
\usepackage{times}
\usepackage[varg]{txfonts}
\usepackage{xcolor}
\usepackage{xspace}
\usepackage{orcidlink}
\usepackage{physics}
\usepackage{graphicx}
\usepackage{dcolumn}
\usepackage{bm,color}
\usepackage{hyperref}
\usepackage{accents}
\usepackage{amssymb,float}
\usepackage{amsmath}
\usepackage{multirow}
\usepackage{siunitx}
\usepackage{tabularx}
\usepackage{subcaption}
\usepackage{booktabs}
\usepackage{url}
\usepackage[shortlabels]{enumitem}

\newcommand{\udt}[3]{#1^{#2}_{\phantom{#2}#3}}

\newcommand{\dut}[3]{#1_{#2}^{\phantom{#2}#3}}

\newcommand{\lc}[1]{\accentset{\circ}{#1}}

\begin{document}

\title{Testing $f(T)$ cosmologies with HII Hubble diagram and CMB distance priors}


\author{Rodrigo Sandoval-Orozco}
\email{rodrigo.sandoval@correo.nucleares.unam.mx}
\affiliation{Instituto de Ciencias Nucleares, Universidad Nacional Aut\'{o}noma de M\'{e}xico, 
Circuito Exterior C.U., A.P. 70-543, M\'exico D.F. 04510, M\'{e}xico.}

\author{Celia Escamilla-Rivera\orcidlink{0000-0002-8929-250X}}
\email{celia.escamilla@nucleares.unam.mx}
\affiliation{Instituto de Ciencias Nucleares, Universidad Nacional Aut\'{o}noma de M\'{e}xico, 
Circuito Exterior C.U., A.P. 70-543, M\'exico D.F. 04510, M\'{e}xico.}

\author{Rebecca Briffa}
\email{rebecca.briffa.16@um.edu.mt}
\affiliation{Institute of Space Sciences and Astronomy, University of Malta, Msida, Malta}
\affiliation{Department of Physics, University of Malta, Msida, Malta}

\author{Jackson Levi Said\orcidlink{0000-0002-7835-4365}}
\email{jackson.said@um.edu.mt}
\affiliation{Institute of Space Sciences and Astronomy, University of Malta, Msida, Malta}
\affiliation{Department of Physics, University of Malta, Msida, Malta}

\begin{abstract}
In this work, we present independent determinations of cosmological parameters and new constraints on $f(T)$ cosmologies, employing two new catalogs related to HII galaxy Hubble and CMB distance priors, along with the local standard measurements, SNIa, $H(z)$ measurements, growth rate data (RSD), and BAO baselines. We found that the marginalised best-fit C.L. $H_0$ and $\sigma_8$ parameters within these cosmologies can relax the current cosmological tensions using HIIG data. This produces a larger range of admissible values for the current Hubble constant, and when all baselines are considered, the uncertainty bands for $H_0$ and the matter density parameter reduce significantly.
\end{abstract}

\maketitle

\section{\label{sec:intro}Introduction}

Several cosmological measurements indicating a cosmic accelerated expansion have suggested that standard cosmology 
itself cannot explain all the observational evidence given by the wide range of surveys and catalogs. Within these collections, observational data from Supernovae Type Ia (SNIa) \cite{SupernovaSearchTeam:1998fmf,SupernovaCosmologyProject:1998vns,Pan-STARRS1:2017jku}, Baryon Acoustic Oscillations (BAO) \cite{Addison:2013haa,Aubourg:2014yra,Cuesta:2014asa,Cuceu:2019for}, and the Cosmic Microwave Background Radiation (CMBR) \cite{Planck:2018vyg} provided a piece of strong evidence that suggests that modifications, in gravitational or matter sectors, needs to be made. Therefore, the easiest modification is to consider a positive cosmological constant $\Lambda$, into the gravitational field equations described by General Relativity (GR) \cite{SupernovaCosmologyProject:1998vns,Weinberg:1988cp,Copeland:2006wr,Frieman:2008sn}, this is the so-called $\Lambda$-Cold Dark Matter ($\Lambda$CDM) concordance model, which is well-constrained using several observational datasets.
However, while this scenario can give us a good description of the current cosmic accelerated expansion, the $\Lambda$CDM model has been facing both theoretical and observational issues~\cite{Verde:2019ivm,Riess:2019qba,DiValentino:2020zio,Riess:2021jrx,DiValentino:2020vvd}. From the theoretical side, so far, we have not reached a clear explanation of the nature of the cosmological constant $\Lambda$, and how the dark energy can be explained by it or whether there are dynamical features to this parameter. Furthermore, the explanation of dark matter is still an open problem within high-energy physics landscapes. On the other hand, on the observational side, there have been careful studies that suggest a revision of the $\Lambda$CDM cosmology since it presents cosmological tensions and discrepancies associated with measured parameter values such as the Hubble constant~\cite{Knox:2019rjx,Jedamzik:2020zmd,DiValentino:2021izs,Abdalla:2022yfr,Kamionkowski:2022pkx,Escudero:2022rbq,Vagnozzi:2023nrq,Khalife:2023qbu}. 

Following this line of thought, one main question arises: 
Can extended theories of gravity be the description of a fundamental theory of gravity that allows us to address theoretical and observational issues with viable solutions in the observed discrepancies?
According to extensive literature, we currently have a global description of cosmological scenarios which can be obtained from the following classifications: \textit{(i)} A standard cosmology with an effective dark energy fluid.
\textit{(ii)} A modified gravity theory in which extra terms due to gravitational effects are related to dark energy effects.
\textit{(iii)} An extension of the teleparallel gravity (TG), where a dark energy fluid can be introduced as an effective fluid \cite{Bahamonde:2021gfp} and references therein, or where the gravitational effects can be related directly to a cosmic late-time acceleration under the non-fluid approach \cite{Aguilar:2024cga}.
These classifications have been successful in giving us an insight into the cosmological tensions, including statistical constraint analysis using SNIa catalogs, Baryon Acoustic Observations (BAO) and $H(z)$ measurements \cite{Briffa:2023ern}, Quasars (QSO) datasets using ultraviolet, X-ray and optical plane techniques \cite{Sandoval-Orozco:2023pit}, and CMB data \cite{Nunes:2018evm,Kumar:2022nvf,Nunes:2018xbm}. However, under the third classification, it has been reported that some deviations can be observed when we consider perturbations of the tetrad field \cite{Aguilar:2024cga}.

As an extension of the role of constraining cosmological models with late-time observational catalogs above mentioned, lensed quasars are interesting objects
since they allow us to measure time delays from gravitational effects. Some of these measurements are available through the H0LiCOW Collaboration \cite{Bonvin:2016crt} and due to
their brightness and variability, we can compute $H_0$
as for standard cosmological scenarios \cite{Kumar:2014vvy,Sereno:2013ona}, for cosmological models derived from modified gravity \cite{DAgostino:2020dhv,Vagnozzi:2023nrq,Dainotti:2021pqg}, and extended cosmologies derived from TG \cite{Sandoval-Orozco:2023pit}, which includes X-ray and optical plane techniques to constraint $H_0$.

Another important measurement in the local universe includes the implementation of HII galaxy Hubble catalogs \citep{Gonzalez-Moran:2021drc}, which have been previously done with observational expansion rates derived using Gaussian Processes (GP) methods in order to optimise values of the model parameters from $f(R)$ models \cite{Sultana:2022qzn}. However, in these analyses when we include spatial curvature it complicates the statistical fitting and leads to high correlations between the free parameters of the models. HII galaxy measurements provide a good source of luminosity distance to probe the cosmic acceleration. One of the interesting methods is the luminosity-velocity dispersion relation $L-\sigma$, which is satisfied by elliptical galaxies, globular clusters and bulges of spiral galaxies. Using this method for the HII observations beyond the current reach of SNIa, we can derive a value for the matter density parameter $\Omega_m$ \cite{Wu:2019mjm}. 
Moreover, while the constraints obtained for this parameter are in good agreement with Planck 2018 data, this result suggests that the dynamical properties of these objects can have a significant impact on the likelihood distributions, especially the ones that constrain the properties of dark energy. 
Further efforts to extend the HII baseline have been investigated in Ref.\cite{Wang:2016pag}, where it is possible to use a combination
of model-dependent and model-independent methods, like GP reconstructions, which give higher values of $H_0$ in comparison to R16 results \cite{Riess:2019cxk}. Further analysis includes constraints on the spatial flatness parameter $\Omega_k$, which avoids biases introduced by a prior model. This analysis was performed using measurements of the Hubble parameter $H(z)$, cosmic chronometers, and its extensions with GP reconstructions of the HII galaxy Hubble diagram \cite{Yennapureddy:2017vvb,Ruan:2019icc}. From astrophysical methodologies, in \cite{Gonzalez-Moran:2021drc} was considered a correlation between the distance estimator and Balmer line luminosity of HII galaxies. This allows us to obtain, in an independent way, constraints on the cosmological parameters derived from standard and modified cosmologies e.g. $H_0$, which are in good agreement with those of similar analyses using SNIa.

At early times, also there have been interesting results regarding extended teleparallel cosmologies, in particular for $f(T)$ cosmologies \cite{Nunes:2018evm,Kumar:2022nvf,Nunes:2018xbm,dosSantos:2021owt,Hashim:2021pkq}. However, the scenarios treated so far, while well-constrained with CMB data, consider the extra torsion-like terms like an effective dark energy fluid in the Boltzmann equations. This later method brings relevant changes in the constraint analysis on cosmological tensions at early times, where the possibility to introduce a non-fluid-like approach as a direct consequence of $f(T)$ extensions has been discussed \cite{Aguilar:2024cga}.
As part of this observational analysis, it is possible to work with a physical and sufficient compression of the full CMB catalog, the so-called CMB distance priors (or shift parameters), which proves to be useful to include CMB data when obtaining cosmological C.L., especially when we are probing dark energy dynamics. The importance of using this kind of catalog is that the CMB shift parameters are the least model-dependent set of parameters that can be extracted from the original CMB catalog, e.g. from Planck 2018 \cite{Wang:2006ts}. In this discussion, there have been studies using this CMB distance priors to understand dark energy dynamics, the primordial power spectrum, and neutrino properties \cite{Zhai:2019nad}, and more recently, using this baseline with multiple tracers method in beyond-standard cosmological probes \cite{Moresco:2022phi}. 

As part of revising and finding new constraints on teleparallel cosmologies, in particular $f(T)$ cosmologies, in this work, we will employ two new catalogs related to HII galaxy Hubble and CMB distance priors, along with the local standard measurements, e.g. SNIa, $H(z)$ measurements, growth rate data (RSD), and BAO baselines. The primordial goal is to find better cosmological constraints for $H_0$ and $\sigma_8$, that allow $f(T)$ cosmologies to relax (or solve) the current cosmological tensions. 

This paper is divided as follows:
In Sec.~\ref{sec:introfT} we summarise the TG background theory and the most promising $f(T)$ cosmologies available in the literature to relax the $H_0$ tension. All of these models are described through their normalised $E(z)$ Friedmann evolution equation. Furthermore, we are going to consider the standard $\Lambda$CDM model in addition to the three $f(T)$ cosmologies described to proceed with comparisons between them.
In Sec.~\ref{sec:obs_data} we present the statistical methodology employed for the baseline datasets mentioned. We divided this discussion into baseline (low-z) local observations and a high-z Hubble diagram.
Our results on new constraints are developed in Sec.~\ref{sec:results-baseline}.
Finally, discussions are presented in Sec.~\ref{sec:conc}.

\section{Teleparallel cosmology background} 
\label{sec:introfT}

Regarding possible extensions to GR, one of the approaches that has been gaining field in cosmological analyses is the one that can be constructed through a metric-affine gravity, which related the connection with other possible geometries \cite{BeltranJimenez:2019esp}. It is along this line of thought that TG offers a curvature-free connection \cite{Bahamonde:2021gfp,Krssak:2018ywd} and produces an interesting formulation of gravity. A novel scenario insight into this approach involves a teleparallel equivalent of general relativity (TEGR), which is dynamically equivalent to GR at the level of 
field equations, and thus
phenomenological predictions, however, it can differ at IR scales \cite{Mylova:2022ljr}.

As discussed in the extensive literature, TG theories are sourced by the interchange of the curvature-based Levi-Civita connection $\udt{\lc{\Gamma}}{\sigma}{\mu\nu}$\footnote{Over-circles notation are used to denote objects determined using the standard Levi-Civita connection.} with the teleparallel connection given by $\udt{\Gamma}{\sigma}{\mu\nu}$ \cite{Hayashi:1979qx,Aldrovandi:2013wha,Bahamonde:2021gfp}. 
Under this scheme, any curvature-less term
of the teleparallel connection means that curvature-based geometries will vanish when we compute with this object, therefore the new objects are needed to construct a set of gravitational theories \cite{Krssak:2018ywd,Aldrovandi:2013wha}. Following this construction, TG can be directly written with the tetrad $\udt{e}{A}{\mu}$ (and its inverses $\dut{E}{A}{\mu}$), and additionally a spin connection $\udt{\omega}{A}{B\mu}$. This tetrad $\udt{e}{A}{\mu}$ is constructed to the metric using the following expressions
\begin{align}\label{metric_tetrad_rel}
    g_{\mu\nu} = \udt{e}{A}{\mu}\udt{e}{B}{\nu}\eta_{AB}\,,& &\eta_{AB} = \dut{E}{A}{\mu}\dut{E}{B}{\nu}g_{\mu\nu}\,.
\end{align}
We will use Latin indices to denote coordinates on the tangent space. The Greek indices denote coordinates on the general manifold. Notice that the tetrad is not the only non-inertial variable in the description of gravity in the standard GR.
Regarding the metric, the tetrad preserves the orthogonality conditions given by:
   $ \udt{e}{A}{\mu}\dut{E}{B}{\mu}=\delta^A_B\,$ and $\udt{e}{A}{\mu}\dut{E}{A}{\nu}=\delta^{\nu}_{\mu}\,.$
Notice that the spin connection $\udt{\omega}{A}{B\mu}$ is a flat object which 
could be responsible for adding the local Lorentz transformation invariance into the equations of evolution. We can join both, the tetrad and spin connection, to define the teleparallel connection as \cite{Weitzenbock1923,Krssak:2018ywd}
\begin{equation}
    \udt{\Gamma}{\sigma}{\nu\mu} \equiv \dut{E}{A}{\sigma}\left(\partial_{\mu}\udt{e}{A}{\nu} + \udt{\omega}{A}{B\mu}\udt{e}{B}{\nu}\right)\,,
\end{equation}
We notice that the spin connection and the tetrad represent the local degrees of freedom and the gravitational terms of the system, and as a consequence, fulfil the diffeomorphism and local Lorentz invariance. 

Furthermore, the torsion tensor can be written from the teleparallel connection given above through the expression \cite{Hayashi:1979qx}
    $\udt{T}{\sigma}{\mu\nu} \equiv 2\udt{\Gamma}{\sigma}{[\nu\mu]}\,,$
where the subindex in square brackets describes the antisymmetric operator. We can compute the torsion scalar by considering the contraction of the torsion tensor as  \cite{Aldrovandi:2013wha,Bahamonde:2021gfp}
\begin{equation}
    T\equiv \frac{1}{4}\udt{T}{\alpha}{\mu\nu}\dut{T}{\alpha}{\mu\nu} + \frac{1}{2}\udt{T}{\alpha}{\mu\nu}\udt{T}{\nu\mu}{\alpha} - \udt{T}{\alpha}{\mu\alpha}\udt{T}{\beta\mu}{\beta}\,.
\end{equation}
An equivalent action, so-called TEGR, is denoted by a linear Lagrangian form of the torsion scalar 
   $ R=\lc{R} + T - B = 0\,,$
\cite{Bahamonde:2015zma,Farrugia:2016qqe},
where $R\equiv0$ denotes curvature-less. $\lc{R} \neq 0$ due to the Levi-Civita connection, and the boundary term $B$ is a total divergence term. Notice that the Einstein-Hilbert action is dynamically equivalent to the description of a linear torsion scalar, therefore it guarantees alike equations of motion for both actions. Modification of TEGR can be written directly with an arbitrary generalisation of the TEGR Lagrangian to the so-called $f(T)$ gravity. In such case, we can parametrise this function of the torsion as $f(T) = -T + \mathcal{F}(T)$, and the action can be written as \cite{Ferraro:2006jd,Ferraro:2008ey,Bengochea:2008gz,Linder:2010py,Chen:2010va,RezaeiAkbarieh:2018ijw} 
\begin{equation}\label{f_T_ext_Lagran}
    \mathcal{S}_{\mathcal{F}(T)}^{} =  \frac{1}{2\kappa^2}\int \mathrm{d}^4 x\; e\left[-T + \mathcal{F}(T)\right] + \int \mathrm{d}^4 x\; e\mathcal{L}_{\text{m}}\,,
\end{equation}
with $\kappa^2=8\pi G$, $\mathcal{L}_{\text{m}}$ is the Lagrangian associated with matter effects, and $e=\det\left(\udt{e}{a}{\mu}\right)=\sqrt{-g}$ is the tetrad determinant. 
It is important to mention that when $\mathcal{F}(T) \rightarrow 0$, the standard $\Lambda$CDM model is recovered when this functional of the torsion tends to a constant $\Lambda$ value. 
Calculating the standard variation of the above action, we obtain the field equations
\begin{equation}
    \dut{S}{\rho}{\mu\nu} = \udt{K}{\mu\nu}{\rho}-\delta_{\rho}^{\mu}T_{\sigma}{}^{\sigma\nu}+\delta_{\rho}^{\nu}T_{\sigma}{}^{\sigma\mu}=-S_{\rho}{}^{\nu\mu}\,.
\end{equation}
where 
\begin{equation}
    \udt{K}{\rho}{\mu\nu} :=\Gamma^{\rho}{}_{\mu\nu}-\lc{\Gamma}^{\rho}{}_{\mu\nu}=\frac{1}{2}\left(T_{\mu}{}^{\rho}{}_{\nu}+T_{\nu}{}^{\rho}{}_{\mu}-T^{\rho}{}_{\mu\nu}\right)\,,
\end{equation}
is the contortion tensor denoting the difference between the Levi-Civi and teleparallel connections.
Each tetrad and spin connection field equations are given by
$    W_{(\mu\nu)} = \kappa^2 \Theta_{\mu\nu}\,,$ and $  \quad W_{[\mu\nu]} = 0\,.$
When we consider a specific metric, a unique tetrad-spin connection pair is allowed and it is in agreement with a vanishing spin connection, this is the so-called Weitzenb\"{o}ck gauge \cite{Krssak:2018ywd,Bahamonde:2021gfp}. 

For our study, we consider a flat homogeneous and isotropic cosmology denoted by the following tetrad \cite{Krssak:2015oua,Tamanini:2012hg}
\begin{equation}
    \udt{e}{A}{\mu} = \text{diag}\left(1,\,a(t),\,a(t),\,a(t)\right)\,,
\end{equation}
where $a(t)$ is the scale factor in physical time $t$. This tetrad satisfies the Weitzenb\"{o}ck gauge condition \cite{Hohmann:2019nat}. Using Eq.~\eqref{metric_tetrad_rel}, we can write the flat Friedmann--Lema\^{i}tre--Robertson--Walker (FLRW) metric as \cite{misner1973gravitation}
\begin{equation}\label{FLRW_metric}
     \mathrm{d}s^2 = \mathrm{d}t^2 - a^2(t) \left(\mathrm{d}x^2+\mathrm{d}y^2+\mathrm{d}z^2\right)\,,
\end{equation}
where the Hubble parameter is $H=\dot{a}/a$ with over-dots referring to derivatives with respect to physical time. Under this description, we notice that $T = -6 H^2$ and $B = -6\left(3H^2 + \dot{H}\right)$. Therefore, the $f(T)$ gravity Friedmann equations are given by the following expressions \cite{Bahamonde:2021gfp}
\begin{align}
    H^2 + \frac{T}{3}\mathcal{F}_T - \frac{\mathcal{F}}{6} &= \frac{\kappa^2}{3}\rho\,,\label{eq:Friedmann_1}\\
    \dot{H}\left(1 - \mathcal{F}_T - 2T\mathcal{F}_{TT}\right) &= -\frac{\kappa^2}{2} \left(\rho + p \right)\label{eq:Friedmann_2}\,,
\end{align}
where $\mathcal{F}_T=\partial \mathcal{F}/\partial T$ and  $\mathcal{F}_{TT}=\partial^2 \mathcal{F}/\partial T^2$ and we indicate the energy density and pressure of the total matter sector by $\rho$ and $p$, respectively. These will be the set of equations to analyse through a selection of viable $f(T)$ cosmologies.

\subsection{$f(T)$ scenarios: models and their cosmologies}

In this section, we describe the models to be constrained through the observational baselines. Notice that these models have been studied extensively in the literature \cite{Bengochea:2008gz,Xu:2018npu,Briffa:2023ozo,Briffa:2021nxg,Linder:2010py,Nesseris:2013jea,Bahamonde:2021gfp}, and references therein. However, a few of them have been successful to specific baselines. We will discuss each of them along this argument.

\begin{itemize}
\item \textbf{Power Law Model} -- 
This model was first studied in \cite{Bengochea:2008gz} due to its ability to reproduce the late-time cosmic acceleration behaviour in the $f(T)$ scheme. We can describe it through
\begin{equation}
\label{eq:f1}
    f_1 (T) = p_0 \left(-T\right)^{p_1}\,,
\end{equation}
where  $p_1$ is a constant and $p_0$ can be computed through $p_1$ at current times by using the $
    p_0 = (6H_0^2)^{1-p_1} (1- \Omega_{m} - \Omega_{r})/(1-2p_1) \,$. 
    We can write the Friedmann equation for this model as
\begin{equation}
    E^2(z) = \Omega_{m} \left(1+z\right)^3 + \Omega_{r}\left(1+z\right)^4 + \left(1 - \Omega_{m} - \Omega_{r}\right) E^{2p_1}(z)\,,
\end{equation}
which recovers $\Lambda$CDM model for $p_1 = 0$. 
For $p_1 = 1$, the extra component in the Friedmann equation gives a re-scaled gravitational constant term in the density parameters, i.e. the GR limit. Also, we can obtain an upper bound such that $p_1 < 1$ for an accelerating Universe.

In \citep{Xu:2018npu} using BAO, SNIa, $H(z)$ and scalar perturbations for the CMB, the authors found that $H_0 = 69.4 \pm 0.8$ km s$^{-1}$ Mpc$^{-1}$, $\Omega_m = 0.298 \pm 0.07$ and $p_1 = -0.10^{+0.09}_{-0.07}$, with a Gaussian prior on $H_0$ in concordance with a local high value. The result is a hint of a slight deviation from the $\Lambda$CDM model for the $p_1$ parameter. In \citep{Briffa:2023ozo} the $f_1(T)$ model results in $H_0 = 69.90\pm 0.58$ km s$^{-1}$ Mpc$^{-1}$, $\Omega_m = 0.289^{+0.016}_{-0.018}$ and $p_1 = 0.014^{+0.091}_{-0.125}$, that indicate a value of $p_1$ close to zero consistent with $\Lambda$CDM. Those results are obtained using $H(z)$, SNIa, BAO and Redshift Space Distortion (RSD) data. Using $H(z)$, SNIa and BAO, in \citep{Briffa:2021nxg} $H_0 = 67.1 \pm 1.6$ km s$^{-1}$ Mpc$^{-1}$, $\Omega_m = 0.294 \pm 0.015$ and $p_1 = 0.06 \pm 0.13$ in agreement with $p_1 = 0$. In \citep{Sandoval-Orozco:2023pit} the combination between $H(z)$, BAO, SNIa and Quasars results in $H_0 = 67.8^{+1.0}_{-0.9}$ km s$^{-1}$ Mpc$^{-1}$, $\Omega_m=0.321 \pm 0.012$ and $p_1 = -0.08^{+0.14}_{-0.15}$ that agrees with $p_1 = 0$.

\item \textbf{Linder Model.} -- 
This model was proposed to produce late-time accelerated expansion through \cite{Linder:2010py} 
\begin{equation}\label{eq:f2}
   f_2 (T) =  p_0 T_0 \left(1 - \text{Exp}\left[-p_2\sqrt{T/T_0}\right]\right)\,,
\end{equation}
where $p_2$ is a constant and $T_0 = T\vert_{t=t_0} = -6H_0^2$. Similar to the previous model, $p_0$ is a constant which can be computed from the Friedman equation at present time through $p_0 = (1- \Omega_{m} - \Omega_{r})/[(1+p_2) e^{-p_2} - 1]\,.$ 
The corresponding Friedmann equation for this model can be written as
\begin{equation}
    E^2\left(z\right) = \Omega_{m} \left(1+z\right)^3 + \Omega_{r}\left(1+z\right)^4 + \frac{1 - \Omega_{m} - \Omega_{r}}{(p_2 + 1)e^{-p_2} - 1} \left[\left(1 + p_2 E(z)\right) \text{Exp}\left[-p_2 E(z)\right] - 1\right]\,,
\end{equation}
which reduces to $\Lambda$CDM as $p_2 \rightarrow +\infty$. To obtain numerical stability, we consider the values for $1/p_2$, so $1/p_2 \rightarrow 0^+$.

In \citep{Xu:2018npu} using the CMB, BAO, SNIa and $H(z)$ the Linder model results in $H_0 = 69.6 \pm 0.9$ km s$^{-1}$ Mpc$^{-1}$, $\Omega_m = 0.296 \pm 0.07$ and $1/p_2 = 0.13^{+0.09}_{-0.11}$ that is not consistent with $1/p_2 \to 0$ which returns the $\Lambda$CDM model, therefore reporting an interesting deviation. Again, a Gaussian prior on $H_0$ is used, consistent with local values. In \citep{Briffa:2023ozo} the authors found that using $H(z)$, SNIa, BAO and RSD $H_0 = 69.38^{+0.67}_{-0.68}$ km s$^{-1}$ Mpc$^{-1}$, $\Omega_m = 0.282 \pm 0.011$ and $1/p_2 = 0.275^{+0.083}_{-0.096}$ that seems to be consistent with the cosmological standard model. Using $H(z)$, SNIa, BAO and Quasars, the authors in \citep{Sandoval-Orozco:2023pit} report that $H_0 = 68.0 \pm 0.7$ km s$^{-1}$ Mpc$^{-1}$, $\Omega_m = 0.300^{0.0008}_{-0.00001}$ and $1/p_2 = 0.048^{+0.178}_{-0.046}$ which is in agreement with $1/p_2 \to 0$ that is a confirmation of the standard cosmological model. Meanwhile in \citep{Briffa:2021nxg} $H_0 = 66.9^{+1.5}_{-1.6}$ km s$^{-1}$ Mpc$^{-1}$, $\Omega_m = 0.294 \pm 0.016$ and $1/p_2 = 0.22^{+0.12}_{-0.15}$ using $H(z)$, SNIa and BAO, resulting in a deviation from the $\Lambda$CDM represented in the $1/p_2$ value. 

\item \textbf{Variant Linder Model} -- \texorpdfstring{$f_3(T)$}{} Model.
A variant version of the latter model can be described by \cite{Nesseris:2013jea} 
\begin{equation}\label{eq:f3}
    f_3 (T) = p_0 T_0\left(1 - \text{Exp}\left[-p_3 T/T_0\right]\right)\,,
\end{equation}
where $p_3$ is constant and $p_0$ in this case can be computed as $p_0 = (1-\Omega_{m} - \Omega_{r})/[(1+2p_3)e^{-p_3}-1]$.
The Friedmann equation for this model can be written as
\begin{equation}
    E^2\left(z\right) = \Omega_{m} \left(1+z\right)^3 + \Omega_{r}\left(1+z\right)^4 + \frac{1 - \Omega_{m} - \Omega_{r}}{(1 + 2p_3 )e^{-p_3} - 1} \left[\left(1 + 2p_3 E^2 (z)\right)\text{Exp}\left[-p_3 E^2 (z)\right] - 1\right]\,,
\end{equation}
which goes to $\Lambda$CDM as $p_3 \rightarrow +\infty$ similar to $f_2$CDM. 

In \citep{Xu:2018npu} using BAO, CMB, SNIa and $H(z)$ the Variant Linder model results in $H_0 = 69.5 \pm 0.8$ km s$^{-1}$ Mpc$^{-1}$, $\Omega_m = 0.297 \pm 0.07$ and $1/p_3 = 0.41 \pm 0.31$ that is not consistent with $1/p_3 \to 0$ and therefore a confirmation of $\Lambda$CDM is not present in the analyses. The results are presented using a Gaussian prior on $H_0$. Using BAO, SNIa, $H(z)$ and RSD in \citep{Briffa:2023ozo} is found that $H_0 = 69.54^{+0.64}_{-0.66}$ km s$^{-1}$ Mpc$^{-1}$, $\Omega_m = 0.282 \pm 0.012$ and $1/p_3 = 0.197^{+0.038}_{-0.092}$ which shows a slight deviation from the expected $1/p_3 \to 0$. Previously, in \citep{Briffa:2021nxg} the combination of $H(z)$, SNIa and BAO reported $H_0 = 67.35^{+0.94}_{-0.97}$ km s$^{-1}$ Mpc$^{-1}$, $\Omega_m = 0.292^{+0.012}_{-0.014}$ and $1/p_3 = 0.043^{+0.101}_{-0.026}$ that again is a slight deviation from the $\Lambda$CDM base model. In \citep{Sandoval-Orozco:2023pit} is presented that $H_0 = 67.2^{+0.8}_{-0.7}$ km s$^{-1}$ Mpc$^{-1}$, $\Omega_m = 0.317^{+0.013}_{-0.012}$ and $1/p_3 = 0.071^{0.097}_{-0.070}$ using SNIa, $H(z)$, BAO and a Quasar Sample. This result is in the range of the expected $1/p_3 \to 0$. 

\end{itemize}


\section{\label{sec:obs_data}Observational data treatment}

In this section, we will evaluate each $f(T)$ cosmological model by employing the constraining parameters method through Markov chain Monte Carlo (MCMC) analysis, utilizing the publicly available code \footnote{\href{https://emcee.readthedocs.io/en/stable/}{emcee.readthedocs.io}} for our cosmology. Our analysis will include a baseline dataset for reference, comprising $H(z)$ measurements, Supernovae Type Ia (SNIa) data, BAO measurements, and two sets of quasar data utilizing ultraviolet, X-ray, and optical plane techniques. Expanding our analysis, we will incorporate additional observations: HII galaxy Hubble, Growth rate data and CMB distance priors. These observations will provide further insights into our cosmological models. 


\subsection{Baseline (low-z) local observations}

\begin{itemize}
\item \textbf{$H(z)$ measurements}. We considered the observational measurements of the Hubble parameter through the Cosmic Chronometers (CC) in which the system of galaxies evolves in a cluster without interacting so small differences in redshift can be measured along with age estimates. The galactic spectra are used to obtain $\mathrm{d}t/\mathrm{d}z$ using redshift and the age difference \citep{Moresco:2016mzx}. The resulting sample contains 31 data points up to $z \sim 2$.  

The corresponding $\chi^2_{\mathrm{CC}}$ is given by: 
\begin{equation}
    \chi^2_{\mathrm{CC}} = \Delta H(z_i,\Theta)^T C^{-1}_{H(z)} \Delta H(z_i,\Theta), 
\end{equation}
where $\Delta H(z_i,\Theta) = H(z_i,\Theta) - H_{obs}(z_i)$ and $ C^{-1}_{\mathrm{CC}}$ is the covariance matrix generated given in \citep{Moresco:2020fbm}. 

\item \textbf{Pantheon+ SNIa dataset.} 
We use the 1701 data points provided by the \textit{Pantheon+} sample \cite{Scolnic:2021amr} that measures the apparent distance for 1550 distinct Supernovae Ia (SNIa) events in a redshift range $0.01 < z < 2.3$. The dataset for the Pantheon+ sample provides SNIa magnitudes corrected for the stretch and colour effects along with the maximum brightness, the mass of the host galaxy, and sky position bias, so to obtain a cosmological useful quantity we need to calculate the distance modulus $\mu = m - M$. The Pantheon+ dataset also includes SH0ES measurements of Cepheid distances to determine $H_0$ and $M$ at the same time without the necessity of including a prior in the parameters \cite{Sandoval-Orozco:2023pit}. Hereafter, this data set will be referred to as PN$^+$.

Thus, the $\chi^2_{\text{SN}}$  for the Pantheon+ sample is given by 
\begin{equation}
    \chi^2_{\text{PN$^+$}} = \Delta\mu(z_i,\Theta)^TC^{-1}_{\text{PN$^+$}} \Delta \mu(z_i,\Theta) + \ln\qty(\frac{S}{2\pi}) - \frac{k^2(\Theta)}{S},
\end{equation}
where $C^{-1}_{\text{PN$^+$}}$ is the total covariance matrix for the data, $S$ is the sum of all components of the inverse of the matrix and $k(\Theta) = \Delta\mu(z_i,\Theta)^TC^{-1}_{\text{PN$^+$}} $. To use SH0ES Cepheids host distances, the SNIa, and the residuals will be calculated as \citep{Brout:2022vxf}: 

\begin{equation}
\Delta\mu(z_i,\Theta) =\left\{
    \begin{matrix}
    \mu_{\text{obs}}(z_i) - \mu_i^{\text{Cepheids}} & i \in\text{Cepheid hosts,} \\
     \mu_{\text{obs}}(z_i) - \mu(z_i,\Theta) & \text{ in other case.}
    \end{matrix} \right.
\end{equation}

For this cases, the distance modulus $\mu(z)$ can be calculated as: 
\begin{equation}
    \mu(z_i,\Theta) = 5\log\qty[D_L(z_i,\Theta)] + M, 
\end{equation}
and where $D_L(z_i,\Theta)$ is the luminosity distance given as: 
\begin{equation}
    D_L(z_i,\Theta) = c(1+z_i)\int_0^{z_i} \frac{dz'}{H(z',\Theta)},
\end{equation}
where $c$ is the speed of light and $H(z_i,\Theta)$ is the Hubble parameter. 

\item \textbf{Baryon acoustic oscillation (BAO) data.}
We will consider a joint baryon acoustic oscillation (BAO) data set composed of several independent data points from different sources. This BAO data set includes measurements from the SDSS Main Galaxy Sample at $z_{\mathrm{eff}} = 0.15$ \cite{Ross:2014qpa}, the six-degree Field Galaxy Survey at $z_{\mathrm{eff}} = 0.106$ \cite{2011MNRAS.416.3017B}, and the BOSS DR11 quasar Lyman-alpha measurement at $z_{\mathrm{eff}} = 2.4$ \cite{Bourboux:2017cbm}. We will also use the angular diameter distances and $H(z)$ measurements of the SDSS-IV eBOSS DR14 quasar survey at $z_{\mathrm{eff}} = \{0.98, 1.23, 1.52, 1.94\}$ \cite{Zhao:2018gvb}, along with the SDSS-III BOSS DR12 consensus BAO measurements of the Hubble parameter and the corresponding comoving angular diameter distances at $z_{\mathrm{eff}} = \{0.38, 0.51, 0.61\}$ \cite{Alam:2016hwk}. For these two BAO data sets, we consider the full covariance matrix in our MCMC analyses as there is an overlap in the redshift slices. 

To use the aforementioned datasets we need to compute additional distances as the angular diameter distance: 
\begin{equation}
    D_V(z) = \left[(1+z)^2D_A(z)^2 \frac{c z}{H(z)}\right]^{1/3},
\end{equation}
where $D_A(z)=(1+z)^{-2}D_L(z)$ is the angular diameter distance. An additional quantity is necessary to perform the BAO analyses as it appears $r_{s,\mathrm{fid}}$ and $r_s(z_d) = r_d$. The value of the first one will be fixed according to the mentioned references as the different results correspond to a specific value. The $r_d$ value will be a free parameter in the MCMC analysis. 

The corresponding $\chi^2$ for the BAO data is calculated using
\begin{equation}
\chi^2_{\text{BAO}}(\Theta) = \Delta G(z_i,\Theta)^T C_{\text{BAO}}^{-1}\Delta G(z_i,\Theta)
\end{equation}
where $\Delta G(z_i,\Theta) = G(z_i,\Theta)-G_{\text{obs}}(z_i)$ using $G(z_i,\Theta)$ the corresponding vector for each of the observations used. $C_{\text{BAO}}$ is the covariance matrix of all the considered BAO observations.

\end{itemize}

\subsection{High-z Hubble diagram}

\begin{itemize}

\item \textbf{HIIG and GHIIR Hubble Diagram (High-z HIIG)}

The HII Galaxies or Giant Extragalactic HII regions GHIIR (that will be referred to as HII from now on) are known to satisfy a relation between the Luminosity in the Balmer Line of Hydrogen $L\mathrm{H}\beta$ and the velocity dispersion $\sigma$ \citep{Gonzalez-Moran:2021drc}. This relation is used as a cosmological distance indicator. HII regions are compact mass systems whose luminosity is composed almost entirely from a young burst of stellar formation with strong narrow emission lines that can be studied in detail \citep{Chavez:2014ria}. 
As we mentioned in the Introduction, the $L-\sigma$ relation can be written 
\begin{equation}
    \log{L} = \alpha\log{\sigma} + \beta, 
\end{equation}
that will be used to calculate the observed distance modulus, which can be expressed from the complete set of observables as \citep{Gonzalez-Moran:2019uij}
\begin{equation}
    \mu_{i,\mathrm{obs}} = 2.5(\alpha + \beta \log{\sigma} - \log{f} -40.08),
\end{equation}
where $\sigma$ is the mentioned velocity dispersion of the object. $\alpha$ and $\beta$ are the intercept and slope of the $L-\sigma$ relation and $\log{f}$ is the measured flux around the continuum. The data points are provided with the flux, the Luminosity $L$ and the dispersion $\sigma$. 

The corresponding $\chi^2$ for the HII sample is calculated using: 
\begin{equation}
    \chi^2_{\mathrm{HII}} = -\frac12 \sum_i \frac{ \left[\mu_{i,\mathrm{obs}} - \mu(z_i,\Theta) \right]^2}{\varepsilon^2},
\end{equation}
where $\mu(z_i,\Theta)$ is the theoretical distance modulus. $\varepsilon$ is the error associated with the measurements that include the statistical uncertainties for every mentioned parameter 
\begin{equation}
\varepsilon^2 = 6.25(\varepsilon^2_{\log{f}} + \beta^2\varepsilon^2_{\log{\sigma}} + \varepsilon^2_{\beta}\log{\sigma^2} + \varepsilon_\alpha^2),
\end{equation}
Although $\alpha$ and $\beta$ are free nuisance parameters, the variation between the numerical values of the parameters for the different models is too small to be considered in this analysis. We will employ $\alpha = 33.268 \pm 0.083$ and $\beta = 5.022 \pm 0.058$ consistent with the values reported in \citep{Gonzalez-Moran:2019uij,Gonzalez-Moran:2021drc,Yang:2024epu,Cao:2023eja}

A sample of 181 points is used in this analysis from \citep{Gonzalez-Moran:2021drc}. Due to the bright nature of the star formation burst, HII composes a promising dataset as they reach $z \sim 4$ 
and could help to complete the Hubble diagram at higher redshifts where SNIa is lacking data points.   


\item \textbf{Growth rate data (RSD)}. The growth rate catalog considered in this work is reported in \cite{Alestas:2022gcg}. The catalog consists of measurements of the growth of cosmic structure, so-called Redshift Space Distortion (RSD). In standard scenarios, the peculiar velocity of galaxies causes cosmic high-density regions larger in the line-of-sight direction at small $z$, given by consequence maps of galaxies where distances are measured from spectroscopic redshifts show anisotropic deviations from the true galaxy distribution. From these results, this baseline can be important to test modified or extended theories of gravity. To derive the growth rate $f(z)$, we can use RSD cosmological probes to constrain cosmological models \cite{Lambiase:2018ows, Gonzalez:2016lur, Gupta:2011kw}. However, instead of deriving the growth rate directly, Large Scale Surveys (LSS) report the density-weighted growth rate, $f\sigma_8 \equiv f(z) \sigma_8(z)$. 
This RSD baseline can be used to constrain cosmological parameters, in particular $\sigma_{8,0}$, e.g. the $\chi^2_{\mathrm{min}}$, which is given by 
\begin{equation}
    \chi^2_\mathrm{RSD}  = \Delta Q(z_i,\Theta)^T C_{\text{RSD}}^{-1}\Delta Q(z_i,\Theta)
\end{equation}
where $Q(z_i,\Theta) =(f\sigma_8(z_i,\Theta)_{\mathrm{theo}} - f\sigma{_8{}_\mathrm{obs}}(z_i))$ and $C_{\text{RSD}}^{-1}$ is the inverse covariance matrix assumed to be a diagonal matrix. The total covariance matrix is given by 
\begin{equation}
C_{\text{RSD}} = 
\begin{pmatrix}
    \sigma_1^2 & 0 & 0 & \dots\\
    0 & C_{\text{WiggleZ}} & 0 & \dots \\
    0 & 0 & \dots & \sigma_N^2 \\
\end{pmatrix} \,.
\end{equation}
Notice that $f\sigma_8(z_i,\Theta)_{\mathrm{obs}}$ is derived from the compiled observational baseline described above.


\item \textbf{CMB (Planck 2018) distance priors.}
The additional sample used for the high redshift universe will be the CMB distance priors (referred to only as CMB in the text). This is a dataset created to provide information on the CMB power spectrum without the calculation of the complete set of perturbations \cite{Zhai:2019nad,Chen:2018dbv}. 

The presented quantities for this dataset are the \textit{shift parameter} $R$ that measures the peak spacing of the temperature in the power spectrum, the \textit{acoustic scale} $l_\mathrm{A}$ that measures the temperature in the transverse direction and the combination $\Omega_b h^2$ \cite{Chen:2018dbv}. Those quantities are known as the distance priors and are going to be calculated as: 
\begin{equation}
    l_\mathrm{A} = (1+z_*)\frac{\pi D_A(z_*)}{r_s(z_*)},
\end{equation}
where $z_d$ is the redshift at photon the decoupling epoch. This redshift value will be fixed to $z = 1089$ in concordance with \citep{Planck:2018vyg}. Additionally: 
\begin{equation}
    R = \frac{(1+z_*)D_A(z_*)\sqrt{\Omega_m H_0^2}}{c},
\end{equation}
with $c$ the light speed. Similar to the analyses performed in BAO, $r_s(z_*) = r_*$ will be used as an additional free parameter. We will use the distance priors as $R=1.7502\pm 0.0046,l_\mathrm{A}=301.471\pm 0.090, \Omega_bh^2=0.02236\pm 0.00015$ with the correlation matrix $c_{ij}$ given as
\begin{equation}
    c_{ij} = 
    \begin{pmatrix}
        1 & 0.46 & -0.66 \\
        0.46 & 1 & -0.33 \\
        -0.66 & -0.34 & 1 
    \end{pmatrix},
\end{equation}
that need to be transformed to a covariance matrix $C_{ij}$ using $C_{ij} = c_{ij}\sigma_i\sigma_j$ with the associated errors for every measurement. So, the $\chi^2$ used for the CMB distance priors is
\begin{equation}
    \chi^2_\mathrm{CMB} = \Delta V(\Theta)^T C_{\text{CMB}}^{-1}\Delta V(\Theta),
\end{equation}
where $V(\Theta) = \left[ R(\Theta),l_\mathrm{A}(\Theta),\Omega_bh^2 \right]$ and $V = (R,l_\mathrm{A},\Omega_bh^2)$ with $\Delta V = V(\Theta) - V$. The normalized Hubble parameter also appears as $h = H_0/100$ km/s/Mpc. 

\end{itemize}

\section{Constraint analyses results}
\label{sec:results-baseline}

This section presents the results for the different models under consideration in this analysis. We compare these results with our baseline $\Lambda$CDM model analysis, which we present first. We consider several combinations of data sets that are chosen to expose important nuances between the behaviours of the models against observational constraints. 

\begin{figure}[H]
    \centering
   \includegraphics[width=7.5cm]{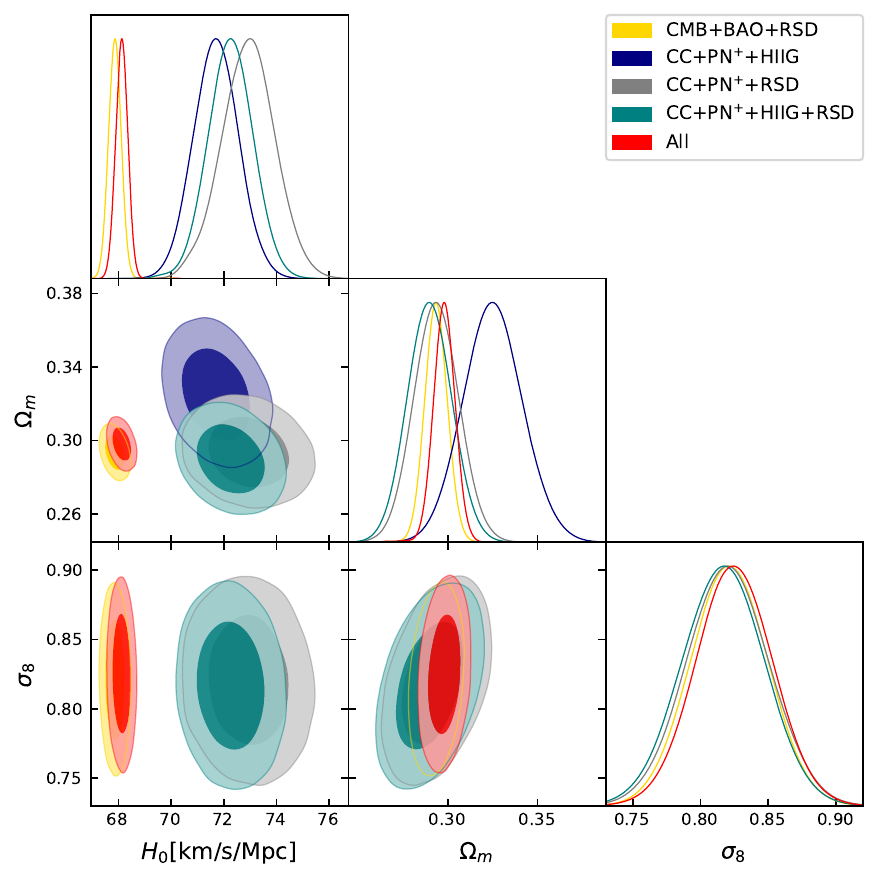}
      \includegraphics[width=7.5cm]{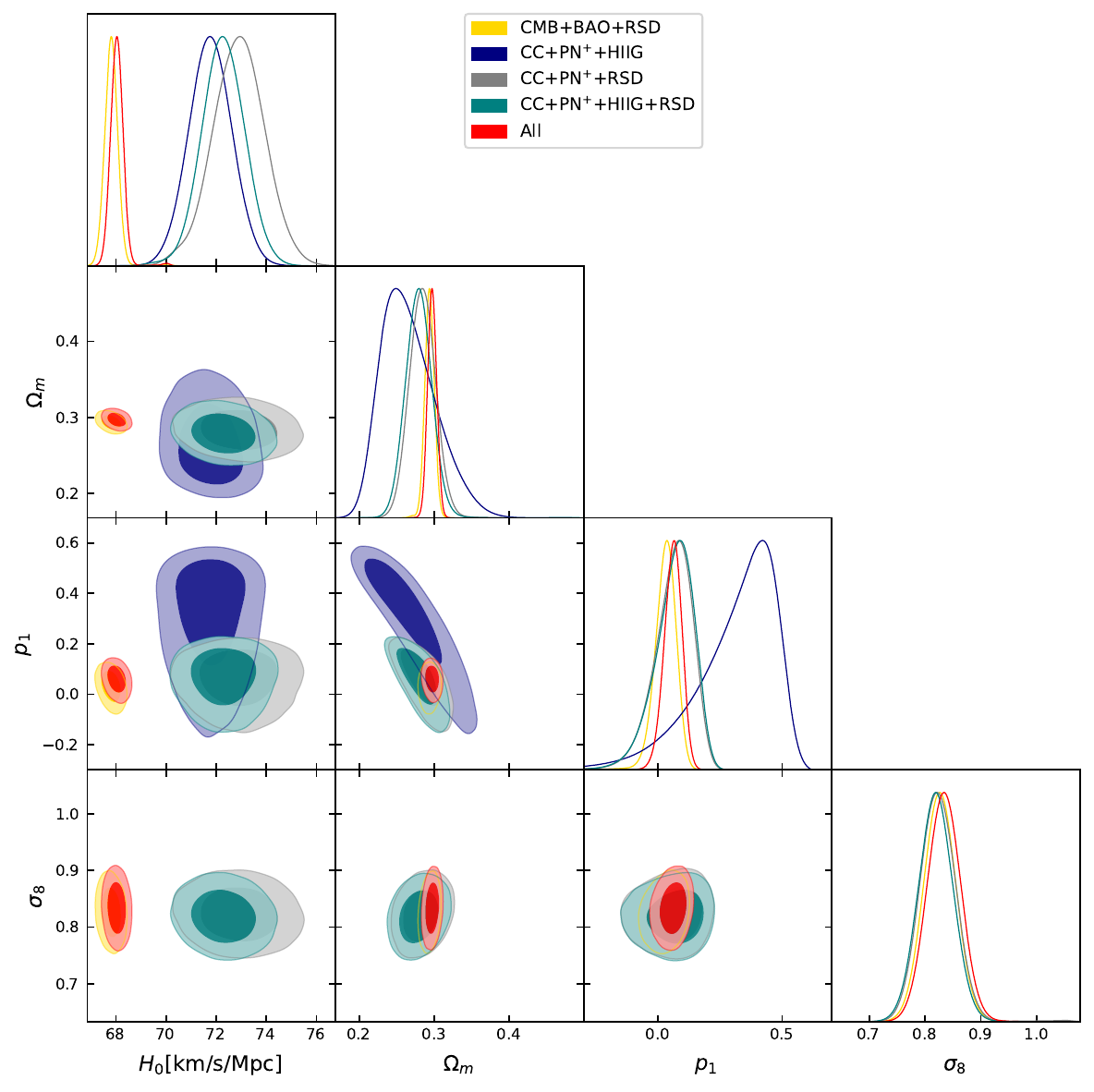}
        \includegraphics[width=7.5cm]{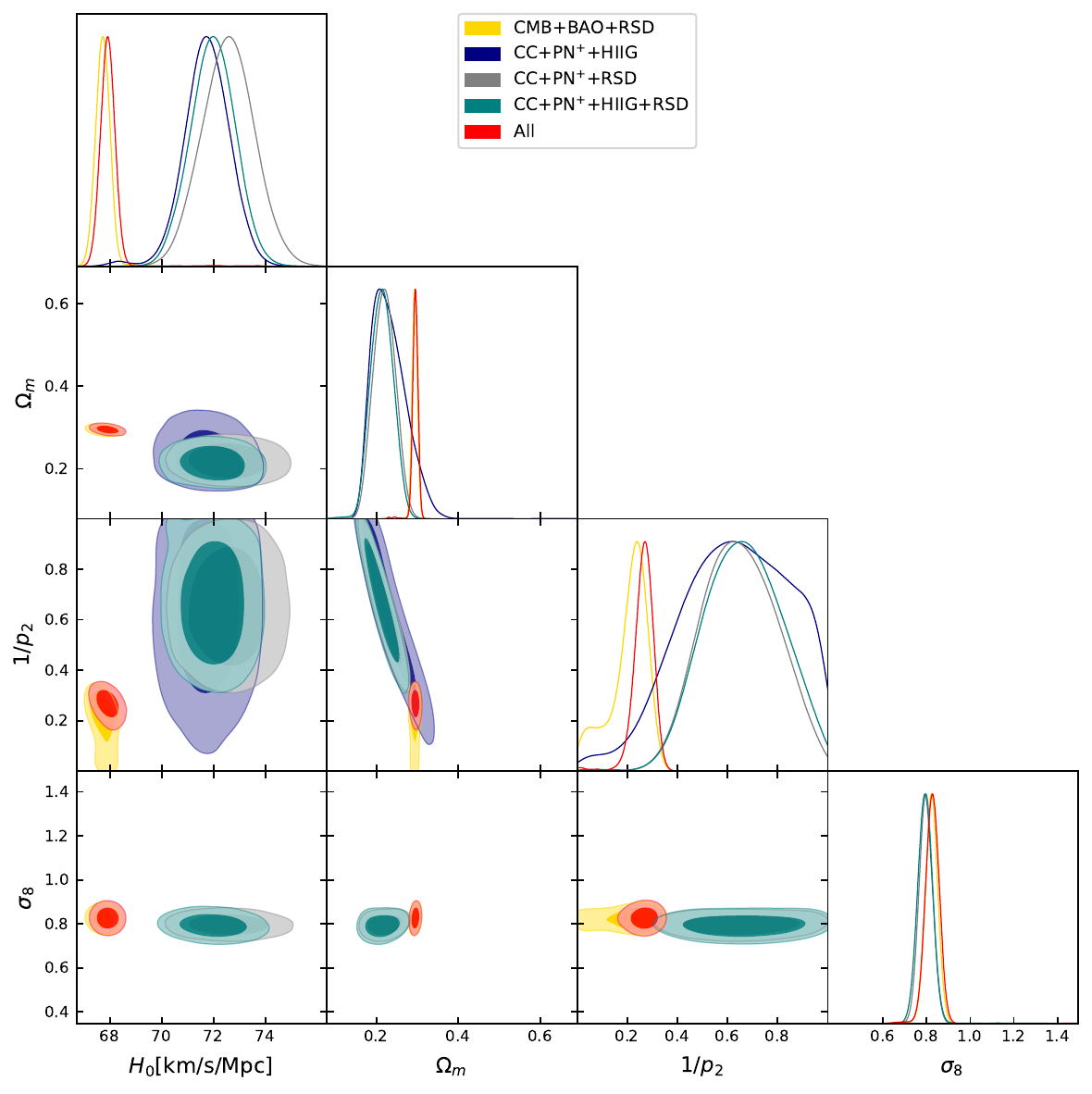}
         \includegraphics[width=7.5cm]{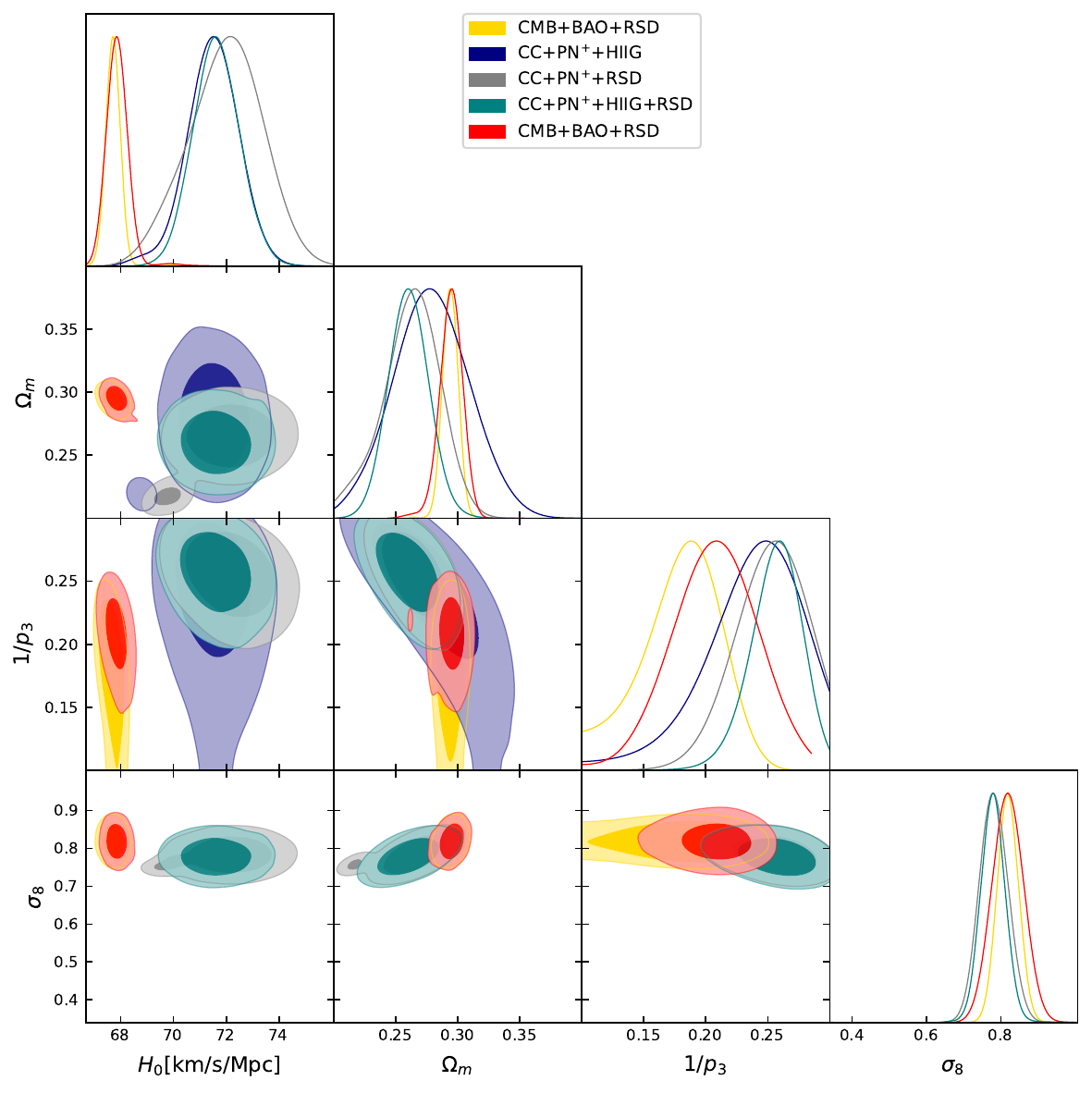}
    \caption{Confidence contours (C.L) and posterior distributions for: \textit{Top left:} the $\Lambda$CDM model parameters, including $H_0$, $\Omega_{m,0}$, and $\sigma_8$. In cases where the RSD data is incorporated (dark blue, green, grey, light blue contours), the $\sigma_{8,0}$ parameter is also displayed. 
    \textit{Top right:} for the $f_1(T)$ model parameters, including $H_0$, $\Omega_{m,0}$, $p_1$, and $\sigma_8$. In cases where the RSD data is incorporated (dark blue, green, grey, light blue contours), the $\sigma_{8,0}$ parameter is also displayed.
    \textit{Bottom left:} for the $f_2(T)$ model parameters, including $H_0$, $\Omega_{m,0}$, $1/p_2$, and $\sigma_8$. In cases where the RSD data is incorporated (dark blue, green, grey, light blue contours), the $\sigma_{8,0}$ parameter is also displayed.
    \textit{Bottom right:} for the $f_3(T)$ model parameters, including $H_0$, $\Omega_{m,0}$ and $1/p_3$. In cases where the RSD data is incorporated (dark blue, green, grey, light blue contours), the $\sigma_{8,0}$ parameter is also displayed.
    }
    \label{fig:lcdm}
\end{figure}

\begin{enumerate}[{(a)}]

\item Constraints for the \textbf{\texorpdfstring{$\Lambda$}{}CDM model}. 
As a baseline, we consider $\Lambda$CDM cosmology with posterior outputs shown in Fig.~\ref{fig:lcdm} with constraints shown in Table~\ref{tab:lcdm}. Immediately, the constraints appear to be strong in that they have robust uncertainties. Also, as is somewhat expected, the PN${}^{+}$ data set appears to push the value of $H_0$ up while the CMB shift data gives a comparatively lower value of the Hubble constant. The differences are smaller when for the matter density parameter $\Omega_m$ and essentially nonexistent for the $\sigma_8$ parameter. One also observes that the absolute magnitude $M$ remains largely consistent throughout the different data sets. While offering a lower number of data points with less certainty, the HIIG data set appears to be largely consistent with the PN${}^{+}$ when considering the value of the Hubble constant. On the other hand, the matter density parameter suffers a significant shift upward in value albeit with a lower level of certainty. This is likely a result of the higher redshift ratio of the data points where the projected Hubble constant value is consistent with PN${}^{+}$ but where more substantial information is obtained on the matter density parameter.

\begin{table*}[htpb!]
\renewcommand{\arraystretch}{0.2}
\resizebox{\textwidth}{!}{%
    \centering
    \begin{tabular}{c |cccccccc}
        \textbf{Dataset} & $H_0$ [km/s/Mpc] & $\Omega_m$ & $M$ & $r_d$ [Mpc] & $r_\ast$ [Mpc] & $\sigma_8$ & $S_8$ (derived) \\
        \hline 
        \hline 
        CC+PN$^{+}$+HII & $71.72 \pm 1.6$ & $0.325^{+0.032}_{-0.030}$ & $-19.292^{+0.047}_{-0.046}$ & -- & -- & -- & -- \\
        CC+PN$^{+}$+RSD & $73.0^{+1.8}_{-2.1}$ & $0.293^{+0.026}_{-0.025}$ & $-19.261^{+0.053}_{-0.063}$ & -- & -- & $0.821\pm 0.060$ &  $0.811^{+0.077}_{-0.075}$ \\
        CC+PN$^{+}$+HIIG+RSD & $72.3^{+1.6}_{-1.6}$ & $0.290^{+0.023}_{-0.023}$ & $-19.287^{+0.047}_{-0.048}$ & -- & -- & $0.817^{+0.057}_{-0.058}$ & $0.803^{+0.076}_{-0.074}$ \\
        CMB+BAO+RSD & $67.90^{+0.46}_{-0.45}$ & $0.290^{+0.010}_{-0.010}$ & -- & $150.1\pm 3.1$ & $148.3 \pm 2.3$ & $0.818^{+0.053}_{-0.054}$ & $0.805^{+0.057}_{-0.058}$ \\
        All & $68.17\pm 0.45$ & $0.293^{+0.011}_{-0.010}$ & $-19.410^{+0.017}_{-0.016}$ & $149.5^{+3.1}_{-3.0}$ & $146.8^{+2.1}_{-2.2}$ & $0.822^{+0.053}_{-0.056}$ & $0.811^{+0.057}_{-0.060}$ \\
        \hline
    \end{tabular}
    }
    \caption{Constraints for the $\Lambda$CDM model that include the parameters $H_0$, $\Omega_{m,0}$ and $\sigma_8$. The $S_8$ parameter (derived) and the nuisance parameter $M$ were computed for each baseline. The full baseline (All) includes CC+PN${}^{+}$+HIIG+RSD+BAO+CMB.}
    \label{tab:lcdm}    
\end{table*}

\item Constraints for the \textbf{Power Law Model -- \texorpdfstring{$f_1(T)$}{} model}.
The power-law model, or $f_1$CDM, is first considered with the constraint table shown in Table~\ref{tab:f1} while the posterior contours are shown in Fig.~\ref{fig:lcdm}. As in the $\Lambda$CDM case, the same division arises between data set combinations involving the PN${}^{+}$ sample and those that do not. As in the $\Lambda$CDM scenario, the Hubble constant is increased for the PN${}^{+}$ sample while the HIIG data set appears to be more consistent with PN${}^{+}$. On the other hand, the HIIG sample allows for a lower value of the matter density parameter as compared with the other data sets. Moreover, there is a clear anti-correlation with the model parameter $p_1$. Indeed, the $p_1$ parameter appears to be counter-balancing the $\Omega_m$ parameter. The constraints on $p_1$ are generally consistent with previous studies \cite{Briffa:2021nxg,Briffa:2023ozo} with a reasonable consistency with its $\Lambda$CDM limit. Another important point to take away from these constraints is how consistent they are at the level of $\sigma_8$, and thus with large-scale structure constraints.

\begin{table}[H]
\resizebox{\textwidth}{!}{%
    \centering
    \begin{tabular}{c|ccccccccc}
        \textbf{Dataset} & $H_0$ [km/s/Mpc] & $\Omega_m$ & $p_1$ & $M$ & $r_d$[Mpc] & $r_\ast$ [Mpc] & $\sigma_8$ & $S_8$ (derived) \\
        \hline 
        \hline 
    CC+PN$^+$+HII & $71.8\pm 1.6$ & $0.267^{+0.073}_{-0.062}$ & $0.31^{+0.25}_{-0.34}$ & $-19.282^{+0.047}_{-0.048}$ & -- & -- & -- & -- \\
        CC+PN$^{+}$+RSD & $72.9^{+2.0}_{-2.1}$ &$0.284^{+0.033}_{-0.032}$ & $0.06^{+0.14}_{-0.16}$ & $-19.266^{+0.056}_{-0.059}$ & -- & -- & $0.824^{+0.061}_{-0.059}$ & $0.802^{+0.085}_{-0.080}$ \\
        CC+PN$^{+}$+HIIG+RSD & $72.3^{+1.6}_{-1.6}$ & $0.280^{+0.033}_{-0.032}$ & $0.06^{+0.14}_{-0.16}$ & $-19.286^{+0.047}_{-0.047}$ & -- & -- & $0.819^{+0.059}_{-0.058}$ & $0.791^{+0.084}_{-0.079}$ \\
        CMB+BAO+RSD & $67.82^{+0.49}_{-0.48}$ & $0.294^{+0.012}_{-0.012}$ & $0.029^{+0.082}_{-0.083}$ & -- & $150.0^{+3.1}_{-3.0}$ & $148.2^{+2.3}_{-2.2}$ & $0.826^{+0.055}_{-0.055}$ & $0.817^{+0.061}_{-0.059}$ \\
        All & $68.05^{+0.47}_{-0.51}$ & $0.297 \pm 0.011$ & $0.059^{+0.069}_{-0.071}$ & $-19.408^{+0.017}_{-0.019}$ & $149.3^{+3.0}_{-3.0}$ & $146.7^{+2.1}_{-2.3}$ & $0.834^{+0.057}_{-0.057}$ & $0.830^{+0.062}_{-0.061}$ \\
        \hline
    \end{tabular}
    }
    \caption{Constraints for the $f_1(T)$ model that include the parameters $H_0$, $\Omega_{m,0}$, $p_1$ and $\sigma_8$. The $S_8$ parameter (derived) and the nuisance parameter $M$ were computed for each baseline. The full baseline (All) includes CC+PN${}^{+}$+HIIG+RSD+BAO+CMB.
    }
    \label{tab:f1}
    
\end{table}

\item Constraints for the \textbf{Linder Model -- \texorpdfstring{$f_2(T)$}{} model}.
The posterior constraints for the $f_2$CDM model are shown in Table~\ref{tab:f2} while their contour fits are in Fig.~\ref{fig:lcdm}. A lot of the nuances that the $f_1$CDM model exhibits are already prevalent in this scenario. On the other hand, the HIIG sample produces a much wider range of admissible values for the Hubble constant, while the value of the matter density parameter is noticeably reduced, while still being consistent with the expected value. Saying that the model parameter $p_2$ remains consistently within the positive range as in previous works \cite{Briffa:2021nxg,Briffa:2023ozo} which also acts as a consistent check on the rest of the analysis. In the perturbative sector, the fits for the $\sigma_8$ parameter are similarly tight and well within the expected range from the plethora of survey releases. 

\begin{table}[H]
\resizebox{\textwidth}{!}{%
    \centering
    \begin{tabular}{c|cccccccc}
        \textbf{Dataset} & $H_0$ [km/s/Mpc] & $\Omega_m$ & $1/p_2$ & $M$ & $r_d$[Mpc] & $r_s$[Mpc] & $\sigma_8$ & $S_8$ (derived) \\
        \hline 
        \hline 
        CC+PN$^+$+HIIG & $71.7^{+1.8}_{-1.7}$ & $0.233^{+0.082}_{-0.071}$ & $0.63^{+0.39}_{-0.39}$ & $-19.280^{+0.052}_{-0.049}$ & -- & -- & -- & -- \\
        CC+PN$^{+}$ + RSD & $72.6^{+1.9}_{-1.9}$ & $0.219^{+0.050}_{-0.048}$ & $0.65^{+0.28}_{-0.26}$ & $-19.258^{+0.056}_{-0.056}$ & -- & -- & $0.796^{+0.059}_{-0.057}$ & $0.680^{+0.10}_{-0.094}$ \\
        CC+PN$^{+}$+HIIG+RSD & $72.0^{+1.6}_{-1.6}$ & $0.213^{+0.050}_{-0.048}$ & $0.66^{+0.28}_{-0.28}$ & $-19.276^{+0.047}_{-0.050}$ & -- & -- & $0.798^{+0.065}_{-0.074}$ & $0.67^{+0.10}_{-0.11}$ \\
        CMB + BAO + RSD &  $67.73^{+0.52}_{-0.51}$ & $0.294^{+0.012}_{-0.012}$ & $0.20^{+0.12}_{-0.20}$ & -- & $149.8^{+3.2}_{-3.1} $ & $148.2^{+2.2}_{-2.2}$ & $0.822^{+0.055}_{-0.056}$ & $0.814^{+0.061}_{-0.062}$ \\
        All & $67.96^{+0.52}_{-0.62}$ & $0.295^{+0.013}_{-0.012}$ & $0.263^{+0.078}_{-0.076}$ & $-19.405^{+0.019}_{-0.023}$ & $148.9^{+3.0}_{-2.9}$ & $147.2^{+2.3}_{-2.3} $  & $0.827^{+0.055}_{-0.060}$ & $0.819^{+0.060}_{-0.065}$ \\
        \hline
    \end{tabular}
    }
    \caption{Constraints for the $f_2(T)$ model that include the parameters $H_0$, $\Omega_{m,0}$ and $\sigma_8$. The $S_8$ parameter (derived) and the nuisance parameter $M$ were computed for each baseline. The full baseline (All) includes CC+PN${}^{+}$+HIIG+RSD+BAO+CMB.}
    \label{tab:f2}
    
\end{table}

\item Constraints for the \textbf{Variant Linder Model -- \texorpdfstring{$f_3(T)$}{} model}. The posteriors and constraint table of this model are shown in Fig.~\ref{fig:lcdm} and Table~\ref{tab:f3}. This is different to the Linder model in that it has a quadratic Hubble dependency in the exponents. In this, the model parameter, which appears in the exponent retains a preference for positive values as in the regular Linder model, albeit with smaller but more confident values. This point is particularly true for the instances where RSD data is used, while the HIIG sample shows a relatively strong preference for a nonzero value for this parameter. On the other hand, and as with the other two models, the CMB, BAO and RSD data push the Hubble constant to lower values while the PN${}^{+}$ data sample prefers higher values of $H_0$. As for the matter density parameter, this can tolerate a lower value which may be more compatible with most independent literature values.

\begin{table}[H]
\resizebox{\textwidth}{!}{
    \centering
    \begin{tabular}{c|cccccccc}
        \textbf{Dataset} & $H_0$ [km/s/Mpc] & $\Omega_m$ & $1/p_3$ & $M$ & $r_d$[Mpc] & $r_*$[Mpc] & $\sigma_8$ & $S_8$ (derived) \\
        \hline 
        \hline 
        CC+PN$^+$+HII & $71.5^{+2.0}_{-2.0}$ & $0.278^{+0.055}_{-0.057}$ & $0.24^{+0.11}_{-0.13}$ & $-19.278^{+0.054}_{-0.049}$ & -- & -- & -- & -- \\
        CC+PN$^{+}$+RSD & $72.0^{+2.1}_{-2.5}$ & $0.261^{+0.032}_{-0.046}$ & $0.264^{+0.12}_{-0.077}$ & $-19.251^{+0.091}_{-0.083}$ & -- & -- & $0.780^{+0.062}_{-0.057}$ & $0.728^{+0.093}_{-0.092}$ \\
        CC+PN$^{+}$+HIIG+RSD & $71.6^{+1.7}_{-1.6}$ & $0.262^{+0.031}_{-0.036}$ & $0.254^{+0.045}_{-0.043}$ & $-19.274^{+0.049}_{-0.048}$ & -- & -- & $0.779^{+0.064}_{-0.062}$ & $0.728^{+0.092}_{-0.097}$ \\
        CMB + BAO + RSD & $67.72^{+0.52}_{-0.52}$ & $0.294\pm 0.012$ & $0.156^{+0.082}_{-0.156}$ & -- & $149.6 \pm 3.2$ & $148.2\pm 2.2$ & $0.819^{+0.055}_{-0.053}$ & $0.810^{+0.060}_{-0.058}$ \\
        All & $67.86^{+0.24}_{-0.26}$ & $0.295\pm 0.005$ & $0.210\pm 0.017$ & $-19.406 \pm 0.011$ & $150.0 \pm 2.5$ & $147.3 \pm 1.3$ & $0.817\pm 0.027$ & $0.809\pm 0.028$ \\
        \hline
    \end{tabular}
    }
    \caption{Constraints for the $f_3(T)$ model that include the parameters $H_0$, $\Omega_{m,0}$ and $\sigma_8$. The $S_8$ parameter (derived) and the nuisance parameter $M$ were computed for each baseline. The full baseline (All) includes CC+PN${}^{+}$+HIIG+RSD+BAO+CMB.}
    \label{tab:f3}
    
\end{table}

\end{enumerate}


\section{\label{sec:conc}Conclusions}

In this study, we have presented the behaviour of parameters across various combinations of datasets. We assessed three prominent $f(T)$ gravity models, evaluating their performance against five distinct combinations of observational datasets. Our combinations included CC+PN$^+$, coupled with either HIIG or RSD datasets, as well as a combined set incorporating all four datasets. Additionally, we explored a dataset combination integrating CMB distance priors and BAO data with RSD data. Finally, we amalgamated all datasets for a holistic analysis.

The inclusion of the RSD dataset proved particularly insightful, as it offers sensitivity to the growth of structure formation, thereby aiding in constraining the $\sigma_{8,0}$ of the models under examination. CMB data was conveniently incorporated to enrich cosmological constraints, and their amalgamation with other cosmic measurements—such as PN$^+$, BAO, and Hubble parameter data—further enhanced our ability to constrain cosmological parameters and alleviate degeneracies between them. Moreover, the utilization of HIIG galaxies, providing luminosity distance measurements, offered further valuable constraints on parameters.

\begin{figure*}
    \centering
    \includegraphics[width=18.0cm]{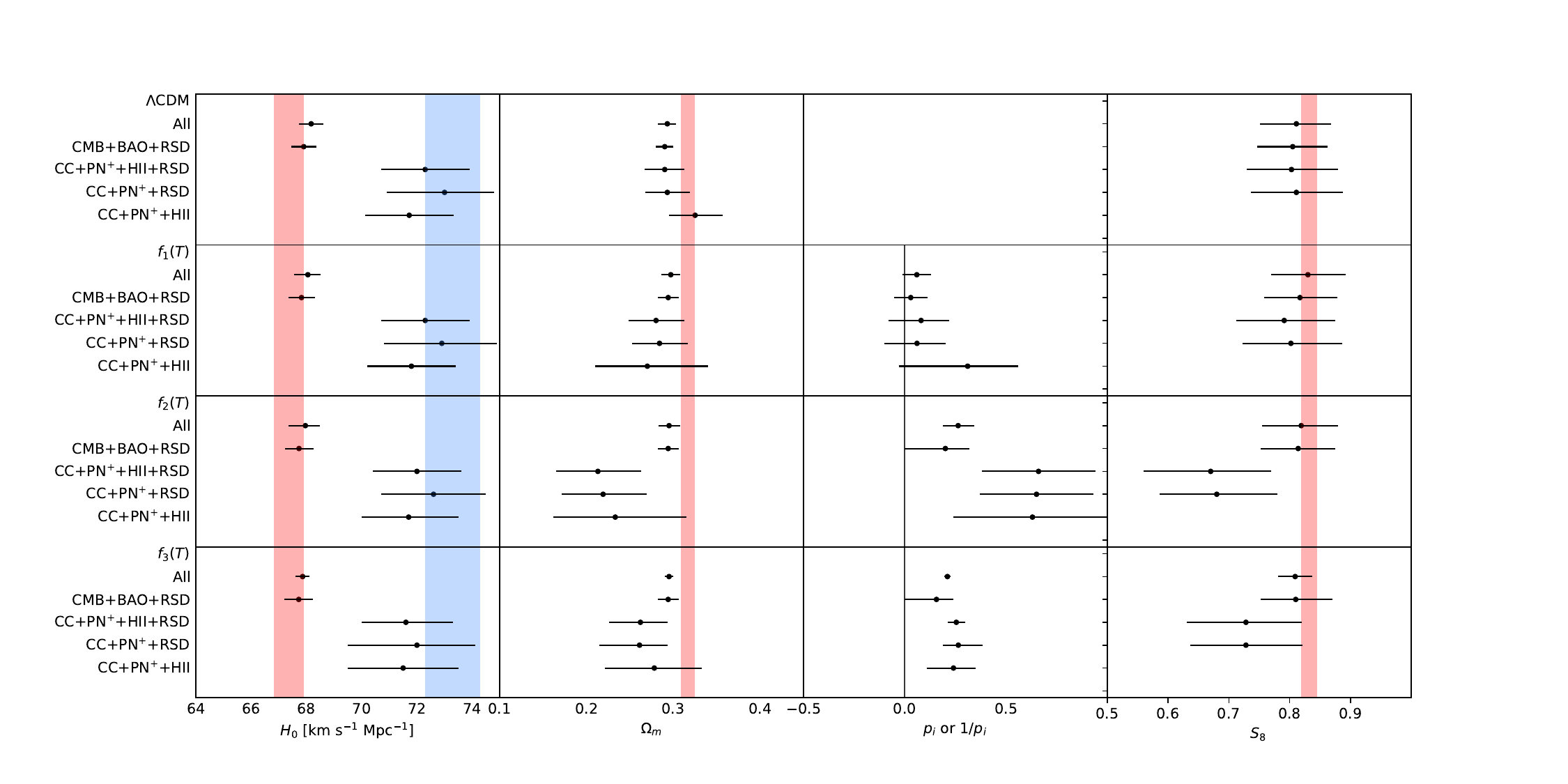}
    \caption{Whisker plot containing the results from all the combinations of baselines and the different $f(T)$ models. The bands in pink and blue colour denote Planck 2018 and SH0ES $H_0$ values respectively.
    }
    \label{fig:whisker}
\end{figure*}

The three primary models under examination are the Power Law model, denoted as $f_1(T)$, the Linder Model, referred to as $f_2(T)$, and a variant of the Linder model, labelled as $f_3(T)$. 
For reference, we also provide the $\Lambda$CDM constraints for each dataset combination considered. The results of the constrained parameters are summarized in Fig.~\ref{fig:whisker}. Notably, when all datasets are included, the uncertainty bands for $H_0$ and the matter density parameter reduce drastically, with the values generally decreasing as well. Additionally, we include values for the derived parameter $S_8$, which remains relatively consistent across various datasets and the three different model types.

Our study provides valuable insights into the behaviour of these $f(T)$ models, particularly in understanding their response to different combinations of observational baseline at higher redshift. It sheds light on their cosmological viability in TG. 
Moving forward, the intention is to use these observations and extend the analysis to encompass the full comprehensive CMB power spectra, along with other early Universe datasets. This broader approach will allow for a more thorough assessment of the competitiveness of these models and their ability to accurately describe the Universe's evolution.

\begin{acknowledgments}
RS is supported by the CONACyT National Grant.
CE-R is supported by the CONACyT Network Project No. 376127 and acknowledges the Royal Astronomical Society as Fellow FRAS 10147. 
This research has been carried out using computational facilities procured through the Cosmostatistics National Group ICN UNAM project.
This article is based upon work from COST Action CA21136 Addressing observational tensions in cosmology with systematics and fundamental physics (CosmoVerse) supported by COST (European Cooperation in Science and Technology). JLS would also like to acknowledge funding from ``The Malta Council for Science and Technology'' as part of the REP-2023-019 (CosmoLearn) Project.
\end{acknowledgments}

\bibliographystyle{JHEP}
\bibliography{references}

\providecommand{\href}[2]{#2}\begingroup\raggedright\begin{thebibliography}{10}

\bibitem{SupernovaSearchTeam:1998fmf}
{\scshape Supernova Search Team} collaboration, \emph{{Observational evidence
  from supernovae for an accelerating universe and a cosmological constant}},
  \href{https://doi.org/10.1086/300499}{\emph{Astron. J.} {\bfseries 116}
  (1998) 1009} [\href{https://arxiv.org/abs/astro-ph/9805201}{{\ttfamily
  astro-ph/9805201}}].

\bibitem{SupernovaCosmologyProject:1998vns}
{\scshape Supernova Cosmology Project} collaboration, \emph{{Measurements of
  $\Omega$ and $\Lambda$ from 42 high redshift supernovae}},
  \href{https://doi.org/10.1086/307221}{\emph{Astrophys. J.} {\bfseries 517}
  (1999) 565} [\href{https://arxiv.org/abs/astro-ph/9812133}{{\ttfamily
  astro-ph/9812133}}].

\bibitem{Pan-STARRS1:2017jku}
{\scshape Pan-STARRS1} collaboration, \emph{{The Complete Light-curve Sample of
  Spectroscopically Confirmed SNe Ia from Pan-STARRS1 and Cosmological
  Constraints from the Combined Pantheon Sample}},
  \href{https://doi.org/10.3847/1538-4357/aab9bb}{\emph{Astrophys. J.}
  {\bfseries 859} (2018) 101}
  [\href{https://arxiv.org/abs/1710.00845}{{\ttfamily 1710.00845}}].

\bibitem{Addison:2013haa}
G.E.~Addison, G.~Hinshaw and M.~Halpern, \emph{{Cosmological constraints from
  baryon acoustic oscillations and clustering of large-scale structure}},
  \href{https://doi.org/10.1093/mnras/stt1687}{\emph{Mon. Not. Roy. Astron.
  Soc.} {\bfseries 436} (2013) 1674}
  [\href{https://arxiv.org/abs/1304.6984}{{\ttfamily 1304.6984}}].

\bibitem{Aubourg:2014yra}
E.~Aubourg et~al., \emph{{Cosmological implications of baryon acoustic
  oscillation measurements}},
  \href{https://doi.org/10.1103/PhysRevD.92.123516}{\emph{Phys. Rev. D}
  {\bfseries 92} (2015) 123516}
  [\href{https://arxiv.org/abs/1411.1074}{{\ttfamily 1411.1074}}].

\bibitem{Cuesta:2014asa}
A.J.~Cuesta, L.~Verde, A.~Riess and R.~Jimenez, \emph{{Calibrating the cosmic
  distance scale ladder: the role of the sound horizon scale and the local
  expansion rate as distance anchors}},
  \href{https://doi.org/10.1093/mnras/stv261}{\emph{Mon. Not. Roy. Astron.
  Soc.} {\bfseries 448} (2015) 3463}
  [\href{https://arxiv.org/abs/1411.1094}{{\ttfamily 1411.1094}}].

\bibitem{Cuceu:2019for}
A.~Cuceu, J.~Farr, P.~Lemos and A.~Font-Ribera, \emph{{Baryon Acoustic
  Oscillations and the Hubble Constant: Past, Present and Future}},
  \href{https://doi.org/10.1088/1475-7516/2019/10/044}{\emph{JCAP} {\bfseries
  10} (2019) 044} [\href{https://arxiv.org/abs/1906.11628}{{\ttfamily
  1906.11628}}].

\bibitem{Planck:2018vyg}
{\scshape Planck} collaboration, \emph{{Planck 2018 results. VI. Cosmological
  parameters}},
  \href{https://doi.org/10.1051/0004-6361/201833910}{\emph{Astron. Astrophys.}
  {\bfseries 641} (2020) A6}
  [\href{https://arxiv.org/abs/1807.06209}{{\ttfamily 1807.06209}}].

\bibitem{Weinberg:1988cp}
S.~Weinberg, \emph{{The Cosmological Constant Problem}},
  \href{https://doi.org/10.1103/RevModPhys.61.1}{\emph{Rev. Mod. Phys.}
  {\bfseries 61} (1989) 1}.

\bibitem{Copeland:2006wr}
E.J.~Copeland, M.~Sami and S.~Tsujikawa, \emph{{Dynamics of dark energy}},
  \href{https://doi.org/10.1142/S021827180600942X}{\emph{Int. J. Mod. Phys. D}
  {\bfseries 15} (2006) 1753}
  [\href{https://arxiv.org/abs/hep-th/0603057}{{\ttfamily hep-th/0603057}}].

\bibitem{Frieman:2008sn}
J.~Frieman, M.~Turner and D.~Huterer, \emph{{Dark Energy and the Accelerating
  Universe}},
  \href{https://doi.org/10.1146/annurev.astro.46.060407.145243}{\emph{Ann. Rev.
  Astron. Astrophys.} {\bfseries 46} (2008) 385}
  [\href{https://arxiv.org/abs/0803.0982}{{\ttfamily 0803.0982}}].

\bibitem{Verde:2019ivm}
L.~Verde, T.~Treu and A.G.~Riess, \emph{{Tensions between the Early and the
  Late Universe}},
  \href{https://doi.org/10.1038/s41550-019-0902-0}{\emph{Nature Astron.}
  {\bfseries 3} (2019) 891} [\href{https://arxiv.org/abs/1907.10625}{{\ttfamily
  1907.10625}}].

\bibitem{Riess:2019qba}
A.G.~Riess, \emph{{The Expansion of the Universe is Faster than Expected}},
  \href{https://doi.org/10.1038/s42254-019-0137-0}{\emph{Nature Rev. Phys.}
  {\bfseries 2} (2019) 10} [\href{https://arxiv.org/abs/2001.03624}{{\ttfamily
  2001.03624}}].

\bibitem{DiValentino:2020zio}
E.~Di~Valentino et~al., \emph{{Snowmass2021 - Letter of interest cosmology
  intertwined II: The hubble constant tension}},
  \href{https://doi.org/10.1016/j.astropartphys.2021.102605}{\emph{Astropart.
  Phys.} {\bfseries 131} (2021) 102605}
  [\href{https://arxiv.org/abs/2008.11284}{{\ttfamily 2008.11284}}].

\bibitem{Riess:2021jrx}
A.G.~Riess et~al., \emph{{A Comprehensive Measurement of the Local Value of the
  Hubble Constant with 1 km s$^{-1}$ Mpc$^{-1}$ Uncertainty from the Hubble
  Space Telescope and the SH0ES Team}},
  \href{https://doi.org/10.3847/2041-8213/ac5c5b}{\emph{Astrophys. J. Lett.}
  {\bfseries 934} (2022) L7}
  [\href{https://arxiv.org/abs/2112.04510}{{\ttfamily 2112.04510}}].

\bibitem{DiValentino:2020vvd}
E.~Di~Valentino et~al., \emph{{Cosmology Intertwined III: $f \sigma_8$ and
  $S_8$}},
  \href{https://doi.org/10.1016/j.astropartphys.2021.102604}{\emph{Astropart.
  Phys.} {\bfseries 131} (2021) 102604}
  [\href{https://arxiv.org/abs/2008.11285}{{\ttfamily 2008.11285}}].

\bibitem{Knox:2019rjx}
L.~Knox and M.~Millea, \emph{{Hubble constant hunter\textquoteright{}s guide}},
  \href{https://doi.org/10.1103/PhysRevD.101.043533}{\emph{Phys. Rev. D}
  {\bfseries 101} (2020) 043533}
  [\href{https://arxiv.org/abs/1908.03663}{{\ttfamily 1908.03663}}].

\bibitem{Jedamzik:2020zmd}
K.~Jedamzik, L.~Pogosian and G.-B.~Zhao, \emph{{Why reducing the cosmic sound
  horizon alone can not fully resolve the Hubble tension}},
  \href{https://doi.org/10.1038/s42005-021-00628-x}{\emph{Commun. in Phys.}
  {\bfseries 4} (2021) 123} [\href{https://arxiv.org/abs/2010.04158}{{\ttfamily
  2010.04158}}].

\bibitem{DiValentino:2021izs}
E.~Di~Valentino, O.~Mena, S.~Pan, L.~Visinelli, W.~Yang, A.~Melchiorri et~al.,
  \emph{{In the realm of the Hubble tension\textemdash{}a review of
  solutions}}, \href{https://doi.org/10.1088/1361-6382/ac086d}{\emph{Class.
  Quant. Grav.} {\bfseries 38} (2021) 153001}
  [\href{https://arxiv.org/abs/2103.01183}{{\ttfamily 2103.01183}}].

\bibitem{Abdalla:2022yfr}
E.~Abdalla et~al., \emph{{Cosmology intertwined: A review of the particle
  physics, astrophysics, and cosmology associated with the cosmological
  tensions and anomalies}},
  \href{https://doi.org/10.1016/j.jheap.2022.04.002}{\emph{JHEAp} {\bfseries
  34} (2022) 49} [\href{https://arxiv.org/abs/2203.06142}{{\ttfamily
  2203.06142}}].

\bibitem{Kamionkowski:2022pkx}
M.~Kamionkowski and A.G.~Riess, \emph{{The Hubble Tension and Early Dark
  Energy}}, {\emph{Ann. Rev. Nucl. Part. Sci.} {\bfseries 73} (2023) 153}
  [\href{https://arxiv.org/abs/2211.04492}{{\ttfamily 2211.04492}}].

\bibitem{Escudero:2022rbq}
H.G.~Escudero, J.-L.~Kuo, R.E.~Keeley and K.N.~Abazajian, \emph{{Early or
  phantom dark energy, self-interacting, extra, or massive neutrinos,
  primordial magnetic fields, or a curved universe: An exploration of possible
  solutions to the H0 and \ensuremath{\sigma}8 problems}},
  \href{https://doi.org/10.1103/PhysRevD.106.103517}{\emph{Phys. Rev. D}
  {\bfseries 106} (2022) 103517}
  [\href{https://arxiv.org/abs/2208.14435}{{\ttfamily 2208.14435}}].

\bibitem{Vagnozzi:2023nrq}
S.~Vagnozzi, \emph{{Seven Hints That Early-Time New Physics Alone Is Not
  Sufficient to Solve the Hubble Tension}},
  \href{https://doi.org/10.3390/universe9090393}{\emph{Universe} {\bfseries 9}
  (2023) 393} [\href{https://arxiv.org/abs/2308.16628}{{\ttfamily
  2308.16628}}].

\bibitem{Khalife:2023qbu}
A.R.~Khalife, M.B.~Zanjani, S.~Galli, S.~G\"unther, J.~Lesgourgues and
  K.~Benabed, \emph{{Review of Hubble tension solutions with new SH0ES and
  SPT-3G data}},  \href{https://arxiv.org/abs/2312.09814}{{\ttfamily
  2312.09814}}.

\bibitem{Bahamonde:2021gfp}
S.~Bahamonde, K.F.~Dialektopoulos, C.~Escamilla-Rivera, G.~Farrugia, V.~Gakis,
  M.~Hendry et~al., \emph{{Teleparallel Gravity: From Theory to Cosmology}},
  \href{https://arxiv.org/abs/2106.13793}{{\ttfamily 2106.13793}}.

\bibitem{Aguilar:2024cga}
A.~Aguilar, C.~Escamilla-Rivera, J.~Levi~Said and J.~Mifsud, \emph{{Non-fluid
  like Boltzmann code architecture for early times f(T) cosmologies}},
  \href{https://arxiv.org/abs/2403.13708}{{\ttfamily 2403.13708}}.

\bibitem{Briffa:2023ern}
R.~Briffa, C.~Escamilla-Rivera, J.~Levi~Said and J.~Mifsud, \emph{{Constraints
  on f(T) cosmology with Pantheon+}},
  \href{https://doi.org/10.1093/mnras/stad1384}{\emph{Mon. Not. Roy. Astron.
  Soc.} {\bfseries 522} (2023) 6024}
  [\href{https://arxiv.org/abs/2303.13840}{{\ttfamily 2303.13840}}].

\bibitem{Sandoval-Orozco:2023pit}
R.~Sandoval-Orozco, C.~Escamilla-Rivera, R.~Briffa and J.~Levi~Said,
  \emph{{f(T) cosmology in the regime of quasar observations}},
  \href{https://doi.org/10.1016/j.dark.2023.101407}{\emph{Phys. Dark Univ.}
  {\bfseries 43} (2024) 101407}
  [\href{https://arxiv.org/abs/2309.03675}{{\ttfamily 2309.03675}}].

\bibitem{Nunes:2018evm}
R.C.~Nunes, S.~Pan and E.N.~Saridakis, \emph{{New observational constraints on
  $f(T)$ gravity through gravitational-wave astronomy}},
  \href{https://doi.org/10.1103/PhysRevD.98.104055}{\emph{Phys. Rev. D}
  {\bfseries 98} (2018) 104055}
  [\href{https://arxiv.org/abs/1810.03942}{{\ttfamily 1810.03942}}].

\bibitem{Kumar:2022nvf}
S.~Kumar, R.C.~Nunes and P.~Yadav, \emph{{New cosmological constraints on f(T)
  gravity in light of full Planck-CMB and type Ia supernovae data}},
  \href{https://doi.org/10.1103/PhysRevD.107.063529}{\emph{Phys. Rev. D}
  {\bfseries 107} (2023) 063529}
  [\href{https://arxiv.org/abs/2209.11131}{{\ttfamily 2209.11131}}].

\bibitem{Nunes:2018xbm}
R.C.~Nunes, \emph{{Structure formation in $f(T)$ gravity and a solution for
  $H_0$ tension}},
  \href{https://doi.org/10.1088/1475-7516/2018/05/052}{\emph{JCAP} {\bfseries
  05} (2018) 052} [\href{https://arxiv.org/abs/1802.02281}{{\ttfamily
  1802.02281}}].

\bibitem{Bonvin:2016crt}
V.~Bonvin et~al., \emph{{H0LiCOW \textendash{} V. New COSMOGRAIL time delays of
  HE 0435\ensuremath{-}1223: $H_0$ to 3.8 per cent precision from strong
  lensing in a flat \ensuremath{\Lambda}CDM model}},
  \href{https://doi.org/10.1093/mnras/stw3006}{\emph{Mon. Not. Roy. Astron.
  Soc.} {\bfseries 465} (2017) 4914}
  [\href{https://arxiv.org/abs/1607.01790}{{\ttfamily 1607.01790}}].

\bibitem{Kumar:2014vvy}
S.R.~Kumar, C.S.~Stalin and T.P.~Prabhu, \emph{{H$_0$ from ten well-measured
  time delay lenses}},
  \href{https://doi.org/10.1051/0004-6361/201423977}{\emph{Astron. Astrophys.}
  {\bfseries 580} (2015) A38}
  [\href{https://arxiv.org/abs/1404.2920}{{\ttfamily 1404.2920}}].

\bibitem{Sereno:2013ona}
M.~Sereno and D.~Paraficz, \emph{{Hubble constant and dark energy inferred from
  free-form determined time delay distances}},
  \href{https://doi.org/10.1093/mnras/stt1938}{\emph{Mon. Not. Roy. Astron.
  Soc.} {\bfseries 437} (2014) 600}
  [\href{https://arxiv.org/abs/1310.2251}{{\ttfamily 1310.2251}}].

\bibitem{DAgostino:2020dhv}
R.~D'Agostino and R.C.~Nunes, \emph{{Measurements of $H_0$ in modified gravity
  theories: The role of lensed quasars in the late-time Universe}},
  \href{https://doi.org/10.1103/PhysRevD.101.103505}{\emph{Phys. Rev. D}
  {\bfseries 101} (2020) 103505}
  [\href{https://arxiv.org/abs/2002.06381}{{\ttfamily 2002.06381}}].

\bibitem{Dainotti:2021pqg}
M.G.~Dainotti, B.~De~Simone, T.~Schiavone, G.~Montani, E.~Rinaldi and
  G.~Lambiase, \emph{{On the Hubble constant tension in the SNe Ia Pantheon
  sample}}, \href{https://doi.org/10.3847/1538-4357/abeb73}{\emph{Astrophys.
  J.} {\bfseries 912} (2021) 150}
  [\href{https://arxiv.org/abs/2103.02117}{{\ttfamily 2103.02117}}].

\bibitem{Gonzalez-Moran:2021drc}
A.L.~Gonz\'alez-Mor\'an, R.~Ch\'avez, E.~Terlevich, R.~Terlevich,
  D.~Fern\'andez-Arenas, F.~Bresolin et~al., \emph{{Independent cosmological
  constraints from high-z H\,ii~galaxies: new results from VLT-KMOS data}},
  \href{https://doi.org/10.1093/mnras/stab1385}{\emph{Mon. Not. Roy. Astron.
  Soc.} {\bfseries 505} (2021) 1441}
  [\href{https://arxiv.org/abs/2105.04025}{{\ttfamily 2105.04025}}].

\bibitem{Sultana:2022qzn}
J.~Sultana, M.K.~Yennapureddy, F.~Melia and D.~Kazanas, \emph{{Constraining
  f(R) models with cosmic chronometers and the H\,ii galaxy Hubble diagram}},
  \href{https://doi.org/10.1093/mnras/stac1713}{\emph{Mon. Not. Roy. Astron.
  Soc.} {\bfseries 514} (2022) 5827}
  [\href{https://arxiv.org/abs/2206.10761}{{\ttfamily 2206.10761}}].

\bibitem{Wu:2019mjm}
Y.~Wu, S.~Cao, J.~Zhang, T.~Liu, Y.~Liu, S.~Geng et~al., \emph{{Exploring the
  ''$L$--$\sigma$'' relation of HII galaxies and giant extragalactic HII
  regions acting as standard candles}},
  \href{https://arxiv.org/abs/1911.10959}{{\ttfamily 1911.10959}}.

\bibitem{Wang:2016pag}
D.~Wang and X.-H.~Meng, \emph{{Determining $H_0$ with the latest HII galaxy
  measurements}},
  \href{https://doi.org/10.3847/1538-4357/aa667e}{\emph{Astrophys. J.}
  {\bfseries 843} (2017) 100}
  [\href{https://arxiv.org/abs/1612.09023}{{\ttfamily 1612.09023}}].

\bibitem{Riess:2019cxk}
A.G.~Riess, S.~Casertano, W.~Yuan, L.M.~Macri and D.~Scolnic, \emph{{Large
  Magellanic Cloud Cepheid Standards Provide a 1\% Foundation for the
  Determination of the Hubble Constant and Stronger Evidence for Physics beyond
  $\Lambda$CDM}},
  \href{https://doi.org/10.3847/1538-4357/ab1422}{\emph{Astrophys. J.}
  {\bfseries 876} (2019) 85}
  [\href{https://arxiv.org/abs/1903.07603}{{\ttfamily 1903.07603}}].

\bibitem{Yennapureddy:2017vvb}
M.K.~Yennapureddy and F.~Melia, \emph{{Reconstruction of the HII Galaxy Hubble
  Diagram using Gaussian Processes}},
  \href{https://doi.org/10.1088/1475-7516/2017/11/029}{\emph{JCAP} {\bfseries
  11} (2017) 029} [\href{https://arxiv.org/abs/1711.03454}{{\ttfamily
  1711.03454}}].

\bibitem{Ruan:2019icc}
C.-Z.~Ruan, F.~Melia, Y.~Chen and T.-J.~Zhang, \emph{{Using spatial curvature
  with HII galaxies and cosmic chronometers to explore the tension in $H_0$}},
  \href{https://doi.org/10.3847/1538-4357/ab2ed0}{\emph{Astrophys. J.}
  {\bfseries 881} (2019) 137}
  [\href{https://arxiv.org/abs/1901.06626}{{\ttfamily 1901.06626}}].

\bibitem{dosSantos:2021owt}
F.B.M.~dos Santos, J.E.~Gonzalez and R.~Silva, \emph{{Observational constraints
  on f(T) gravity from model-independent data}},
  \href{https://doi.org/10.1140/epjc/s10052-022-10784-1}{\emph{Eur. Phys. J. C}
  {\bfseries 82} (2022) 823}
  [\href{https://arxiv.org/abs/2112.15249}{{\ttfamily 2112.15249}}].

\bibitem{Hashim:2021pkq}
M.~Hashim, A.A.~El-Zant, W.~El~Hanafy and A.~Golovnev, \emph{{Toward a
  concordance teleparallel cosmology. Part~II. Linear perturbation}},
  \href{https://doi.org/10.1088/1475-7516/2021/07/053}{\emph{JCAP} {\bfseries
  07} (2021) 053} [\href{https://arxiv.org/abs/2104.08311}{{\ttfamily
  2104.08311}}].

\bibitem{Wang:2006ts}
Y.~Wang and P.~Mukherjee, \emph{{Robust dark energy constraints from
  supernovae, galaxy clustering, and three-year wilkinson microwave anisotropy
  probe observations}}, \href{https://doi.org/10.1086/507091}{\emph{Astrophys.
  J.} {\bfseries 650} (2006) 1}
  [\href{https://arxiv.org/abs/astro-ph/0604051}{{\ttfamily
  astro-ph/0604051}}].

\bibitem{Zhai:2019nad}
Z.~Zhai, C.-G.~Park, Y.~Wang and B.~Ratra, \emph{{CMB distance priors
  revisited: effects of dark energy dynamics, spatial curvature, primordial
  power spectrum, and neutrino parameters}},
  \href{https://doi.org/10.1088/1475-7516/2020/07/009}{\emph{JCAP} {\bfseries
  07} (2020) 009} [\href{https://arxiv.org/abs/1912.04921}{{\ttfamily
  1912.04921}}].

\bibitem{Moresco:2022phi}
M.~Moresco et~al., \emph{{Unveiling the Universe with emerging cosmological
  probes}}, \href{https://doi.org/10.1007/s41114-022-00040-z}{\emph{Living Rev.
  Rel.} {\bfseries 25} (2022) 6}
  [\href{https://arxiv.org/abs/2201.07241}{{\ttfamily 2201.07241}}].

\bibitem{BeltranJimenez:2019esp}
J.~Beltr\'an~Jim\'enez, L.~Heisenberg and T.S.~Koivisto, \emph{{The Geometrical
  Trinity of Gravity}},
  \href{https://doi.org/10.3390/universe5070173}{\emph{Universe} {\bfseries 5}
  (2019) 173} [\href{https://arxiv.org/abs/1903.06830}{{\ttfamily
  1903.06830}}].

\bibitem{Krssak:2018ywd}
M.~Krssak, R.J.~van~den Hoogen, J.G.~Pereira, C.G.~B\"ohmer and A.A.~Coley,
  \emph{{Teleparallel theories of gravity: illuminating a fully invariant
  approach}}, \href{https://doi.org/10.1088/1361-6382/ab2e1f}{\emph{Class.
  Quant. Grav.} {\bfseries 36} (2019) 183001}
  [\href{https://arxiv.org/abs/1810.12932}{{\ttfamily 1810.12932}}].

\bibitem{Mylova:2022ljr}
M.~Mylova, J.~Levi~Said and E.N.~Saridakis, \emph{{General Effective Field
  Theory of Teleparallel Gravity}},
  \href{https://arxiv.org/abs/2211.11420}{{\ttfamily 2211.11420}}.

\bibitem{Hayashi:1979qx}
K.~Hayashi and T.~Shirafuji, \emph{{New General Relativity}},
  \href{https://doi.org/10.1103/PhysRevD.19.3524}{\emph{Phys. Rev. D}
  {\bfseries 19} (1979) 3524}.

\bibitem{Aldrovandi:2013wha}
R.~Aldrovandi and J.G.~Pereira, \emph{{Teleparallel Gravity}: {An
  Introduction}}, Springer (2013),
  \href{https://doi.org/10.1007/978-94-007-5143-9}{10.1007/978-94-007-5143-9}.

\bibitem{Weitzenbock1923}
R.~Weitzenb\"{o}ock, \emph{`Invariantentheorie'}, Noordhoff, Gronningen (1923).

\bibitem{Bahamonde:2015zma}
S.~Bahamonde, C.G.~B\"ohmer and M.~Wright, \emph{{Modified teleparallel
  theories of gravity}},
  \href{https://doi.org/10.1103/PhysRevD.92.104042}{\emph{Phys. Rev. D}
  {\bfseries 92} (2015) 104042}
  [\href{https://arxiv.org/abs/1508.05120}{{\ttfamily 1508.05120}}].

\bibitem{Farrugia:2016qqe}
G.~Farrugia and J.~Levi~Said, \emph{{Stability of the flat FLRW metric in
  $f(T)$ gravity}},
  \href{https://doi.org/10.1103/PhysRevD.94.124054}{\emph{Phys. Rev. D}
  {\bfseries 94} (2016) 124054}
  [\href{https://arxiv.org/abs/1701.00134}{{\ttfamily 1701.00134}}].

\bibitem{Ferraro:2006jd}
R.~Ferraro and F.~Fiorini, \emph{{Modified teleparallel gravity: Inflation
  without inflaton}},
  \href{https://doi.org/10.1103/PhysRevD.75.084031}{\emph{Phys. Rev.}
  {\bfseries D75} (2007) 084031}
  [\href{https://arxiv.org/abs/gr-qc/0610067}{{\ttfamily gr-qc/0610067}}].

\bibitem{Ferraro:2008ey}
R.~Ferraro and F.~Fiorini, \emph{{On Born-Infeld Gravity in Weitzenbock
  spacetime}}, \href{https://doi.org/10.1103/PhysRevD.78.124019}{\emph{Phys.
  Rev.} {\bfseries D78} (2008) 124019}
  [\href{https://arxiv.org/abs/0812.1981}{{\ttfamily 0812.1981}}].

\bibitem{Bengochea:2008gz}
G.R.~Bengochea and R.~Ferraro, \emph{{Dark torsion as the cosmic speed-up}},
  \href{https://doi.org/10.1103/PhysRevD.79.124019}{\emph{Phys. Rev.}
  {\bfseries D79} (2009) 124019}
  [\href{https://arxiv.org/abs/0812.1205}{{\ttfamily 0812.1205}}].

\bibitem{Linder:2010py}
E.V.~Linder, \emph{{Einstein's Other Gravity and the Acceleration of the
  Universe}}, \href{https://doi.org/10.1103/PhysRevD.81.127301,
  10.1103/PhysRevD.82.109902}{\emph{Phys. Rev.} {\bfseries D81} (2010) 127301}
  [\href{https://arxiv.org/abs/1005.3039}{{\ttfamily 1005.3039}}].

\bibitem{Chen:2010va}
S.-H.~Chen, J.B.~Dent, S.~Dutta and E.N.~Saridakis, \emph{{Cosmological
  perturbations in f(T) gravity}},
  \href{https://doi.org/10.1103/PhysRevD.83.023508}{\emph{Phys. Rev.}
  {\bfseries D83} (2011) 023508}
  [\href{https://arxiv.org/abs/1008.1250}{{\ttfamily 1008.1250}}].

\bibitem{RezaeiAkbarieh:2018ijw}
A.~Rezaei~Akbarieh and Y.~Izadi, \emph{{Tachyon Inflation in Teleparallel
  Gravity}}, \href{https://doi.org/10.1140/epjc/s10052-019-6819-z}{\emph{Eur.
  Phys. J. C} {\bfseries 79} (2019) 366}
  [\href{https://arxiv.org/abs/1812.06649}{{\ttfamily 1812.06649}}].

\bibitem{Krssak:2015oua}
M.~Kr\v{s}\v{s}\'ak and E.N.~Saridakis, \emph{{The covariant formulation of
  f(T) gravity}},
  \href{https://doi.org/10.1088/0264-9381/33/11/115009}{\emph{Class. Quant.
  Grav.} {\bfseries 33} (2016) 115009}
  [\href{https://arxiv.org/abs/1510.08432}{{\ttfamily 1510.08432}}].

\bibitem{Tamanini:2012hg}
N.~Tamanini and C.G.~Boehmer, \emph{{Good and bad tetrads in f(T) gravity}},
  \href{https://doi.org/10.1103/PhysRevD.86.044009}{\emph{Phys. Rev. D}
  {\bfseries 86} (2012) 044009}
  [\href{https://arxiv.org/abs/1204.4593}{{\ttfamily 1204.4593}}].

\bibitem{Hohmann:2019nat}
M.~Hohmann, L.~J\"arv, M.~Kr\v{s}\v{s}\'ak and C.~Pfeifer, \emph{{Modified
  teleparallel theories of gravity in symmetric spacetimes}},
  \href{https://doi.org/10.1103/PhysRevD.100.084002}{\emph{Phys. Rev. D}
  {\bfseries 100} (2019) 084002}
  [\href{https://arxiv.org/abs/1901.05472}{{\ttfamily 1901.05472}}].

\bibitem{misner1973gravitation}
C.~Misner, K.~Thorne and J.~Wheeler, \emph{Gravitation}, no.~pt. 3 in
  Gravitation, W. H. Freeman (1973).

\bibitem{Xu:2018npu}
B.~Xu, H.~Yu and P.~Wu, \emph{{Testing Viable f(T) Models with Current
  Observations}},
  \href{https://doi.org/10.3847/1538-4357/aaad12}{\emph{Astrophys. J.}
  {\bfseries 855} (2018) 89}.

\bibitem{Briffa:2023ozo}
R.~Briffa, C.~Escamilla-Rivera, J.~Levi~Said and J.~Mifsud, \emph{{Growth of
  structures using redshift space distortion in f(T) cosmology}},
  \href{https://doi.org/10.1093/mnras/stae103}{\emph{Mon. Not. Roy. Astron.
  Soc.} {\bfseries 528} (2024) 2711}
  [\href{https://arxiv.org/abs/2310.09159}{{\ttfamily 2310.09159}}].

\bibitem{Briffa:2021nxg}
R.~Briffa, C.~Escamilla-Rivera, J.~Said~Levi, J.~Mifsud and N.L.~Pullicino,
  \emph{{Impact of $H_0$ priors on $f(T)$ late time cosmology}},
  \href{https://doi.org/10.1140/epjp/s13360-022-02725-4}{\emph{Eur. Phys. J.
  Plus} {\bfseries 137} (2022) 532}
  [\href{https://arxiv.org/abs/2108.03853}{{\ttfamily 2108.03853}}].

\bibitem{Nesseris:2013jea}
S.~Nesseris, S.~Basilakos, E.N.~Saridakis and L.~Perivolaropoulos,
  \emph{{Viable $f(T)$ models are practically indistinguishable from
  $\Lambda$CDM}}, \href{https://doi.org/10.1103/PhysRevD.88.103010}{\emph{Phys.
  Rev. D} {\bfseries 88} (2013) 103010}
  [\href{https://arxiv.org/abs/1308.6142}{{\ttfamily 1308.6142}}].

\bibitem{Moresco:2016mzx}
M.~Moresco, L.~Pozzetti, A.~Cimatti, R.~Jimenez, C.~Maraston, L.~Verde et~al.,
  \emph{{A 6\% measurement of the Hubble parameter at $z\sim0.45$: direct
  evidence of the epoch of cosmic re-acceleration}},
  \href{https://doi.org/10.1088/1475-7516/2016/05/014}{\emph{JCAP} {\bfseries
  05} (2016) 014} [\href{https://arxiv.org/abs/1601.01701}{{\ttfamily
  1601.01701}}].

\bibitem{Moresco:2020fbm}
M.~Moresco, R.~Jimenez, L.~Verde, A.~Cimatti and L.~Pozzetti, \emph{{Setting
  the Stage for Cosmic Chronometers. II. Impact of Stellar Population Synthesis
  Models Systematics and Full Covariance Matrix}},
  \href{https://doi.org/10.3847/1538-4357/ab9eb0}{\emph{Astrophys. J.}
  {\bfseries 898} (2020) 82}
  [\href{https://arxiv.org/abs/2003.07362}{{\ttfamily 2003.07362}}].

\bibitem{Scolnic:2021amr}
D.~Scolnic et~al., \emph{{The Pantheon+ Analysis: The Full Data Set and
  Light-curve Release}},
  \href{https://doi.org/10.3847/1538-4357/ac8b7a}{\emph{Astrophys. J.}
  {\bfseries 938} (2022) 113}
  [\href{https://arxiv.org/abs/2112.03863}{{\ttfamily 2112.03863}}].

\bibitem{Brout:2022vxf}
D.~Brout et~al., \emph{{The Pantheon+ Analysis: Cosmological Constraints}},
  \href{https://doi.org/10.3847/1538-4357/ac8e04}{\emph{Astrophys. J.}
  {\bfseries 938} (2022) 110}
  [\href{https://arxiv.org/abs/2202.04077}{{\ttfamily 2202.04077}}].

\bibitem{Ross:2014qpa}
A.J.~Ross, L.~Samushia, C.~Howlett, W.J.~Percival, A.~Burden and M.~Manera,
  \emph{{The clustering of the SDSS DR7 main Galaxy sample \textendash{} I. A 4
  per cent distance measure at $z = 0.15$}},
  \href{https://doi.org/10.1093/mnras/stv154}{\emph{Mon. Not. Roy. Astron.
  Soc.} {\bfseries 449} (2015) 835}
  [\href{https://arxiv.org/abs/1409.3242}{{\ttfamily 1409.3242}}].

\bibitem{2011MNRAS.416.3017B}
F.~{Beutler}, C.~{Blake}, M.~{Colless}, D.H.~{Jones}, L.~{Staveley-Smith},
  L.~{Campbell} et~al., \emph{{The 6dF Galaxy Survey: baryon acoustic
  oscillations and the local Hubble constant}},
  \href{https://doi.org/10.1111/j.1365-2966.2011.19250.x}{\emph{Monthly Notices
  of the Royal Astronomical Society} {\bfseries 416} (2011) 3017}
  [\href{https://arxiv.org/abs/1106.3366}{{\ttfamily 1106.3366}}].

\bibitem{Bourboux:2017cbm}
H.~du~Mas~des Bourboux et~al., \emph{{Baryon acoustic oscillations from the
  complete SDSS-III Ly$\alpha$-quasar cross-correlation function at $z=2.4$}},
  \href{https://doi.org/10.1051/0004-6361/201731731}{\emph{Astron. Astrophys.}
  {\bfseries 608} (2017) A130}
  [\href{https://arxiv.org/abs/1708.02225}{{\ttfamily 1708.02225}}].

\bibitem{Zhao:2018gvb}
G.-B.~Zhao et~al., \emph{{The clustering of the SDSS-IV extended Baryon
  Oscillation Spectroscopic Survey DR14 quasar sample: a tomographic
  measurement of cosmic structure growth and expansion rate based on optimal
  redshift weights}}, \href{https://doi.org/10.1093/mnras/sty2845}{\emph{Mon.
  Not. Roy. Astron. Soc.} {\bfseries 482} (2019) 3497}
  [\href{https://arxiv.org/abs/1801.03043}{{\ttfamily 1801.03043}}].

\bibitem{Alam:2016hwk}
{\scshape BOSS} collaboration, \emph{{The clustering of galaxies in the
  completed SDSS-III Baryon Oscillation Spectroscopic Survey: cosmological
  analysis of the DR12 galaxy sample}},
  \href{https://doi.org/10.1093/mnras/stx721}{\emph{Mon. Not. Roy. Astron.
  Soc.} {\bfseries 470} (2017) 2617}
  [\href{https://arxiv.org/abs/1607.03155}{{\ttfamily 1607.03155}}].

\bibitem{Chavez:2014ria}
R.~Ch\'avez, R.~Terlevich, E.~Terlevich, F.~Bresolin, J.~Melnick, M.~Plionis
  et~al., \emph{{The L\textendash{}\ensuremath{\sigma} relation for massive
  bursts of star formation}},
  \href{https://doi.org/10.1093/mnras/stu987}{\emph{Mon. Not. Roy. Astron.
  Soc.} {\bfseries 442} (2014) 3565}
  [\href{https://arxiv.org/abs/1405.4010}{{\ttfamily 1405.4010}}].

\bibitem{Gonzalez-Moran:2019uij}
A.L.~Gonz\'alez-Mor\'an, R.~Ch\'avez, R.~Terlevich, E.~Terlevich, F.~Bresolin,
  D.~Fern\'andez-Arenas et~al., \emph{{Independent cosmological constraints
  from high-z H ii galaxies}},
  \href{https://doi.org/10.1093/mnras/stz1577}{\emph{Mon. Not. Roy. Astron.
  Soc.} {\bfseries 487} (2019) 4669}
  [\href{https://arxiv.org/abs/1906.02195}{{\ttfamily 1906.02195}}].

\bibitem{Yang:2024epu}
Y.~Yang, T.~Liu, J.~Huang, X.~Cheng, M.~Biesiada and S.-m.~Wu,
  \emph{{Simultaneous measurements on cosmic curvature and opacity using latest
  HII regions and H(z) observations}},
  \href{https://doi.org/10.1140/epjc/s10052-023-12356-3}{\emph{Eur. Phys. J. C}
  {\bfseries 84} (2024) 3} [\href{https://arxiv.org/abs/2401.03413}{{\ttfamily
  2401.03413}}].

\bibitem{Cao:2023eja}
S.~Cao and B.~Ratra, \emph{{H0=69.8\ensuremath{\pm}1.3\,\,km\,s-1\,Mpc-1,
  \ensuremath{\Omega}m0=0.288\ensuremath{\pm}0.017, and other constraints from
  lower-redshift, non-CMB, expansion-rate data}},
  \href{https://doi.org/10.1103/PhysRevD.107.103521}{\emph{Phys. Rev. D}
  {\bfseries 107} (2023) 103521}
  [\href{https://arxiv.org/abs/2302.14203}{{\ttfamily 2302.14203}}].

\bibitem{Alestas:2022gcg}
G.~Alestas, L.~Kazantzidis and S.~Nesseris, \emph{{Machine learning constraints
  on deviations from general relativity from the large scale structure of the
  Universe}}, \href{https://doi.org/10.1103/PhysRevD.106.103519}{\emph{Phys.
  Rev. D} {\bfseries 106} (2022) 103519}
  [\href{https://arxiv.org/abs/2209.12799}{{\ttfamily 2209.12799}}].

\bibitem{Lambiase:2018ows}
G.~Lambiase, S.~Mohanty, A.~Narang and P.~Parashari, \emph{{Testing dark energy
  models in the light of $\sigma _8$ tension}},
  \href{https://doi.org/10.1140/epjc/s10052-019-6634-6}{\emph{Eur. Phys. J. C}
  {\bfseries 79} (2019) 141}
  [\href{https://arxiv.org/abs/1804.07154}{{\ttfamily 1804.07154}}].

\bibitem{Gonzalez:2016lur}
J.E.~Gonzalez, J.S.~Alcaniz and J.C.~Carvalho, \emph{{Non-parametric
  reconstruction of cosmological matter perturbations}},
  \href{https://doi.org/10.1088/1475-7516/2016/04/016}{\emph{JCAP} {\bfseries
  04} (2016) 016} [\href{https://arxiv.org/abs/1602.01015}{{\ttfamily
  1602.01015}}].

\bibitem{Gupta:2011kw}
G.~Gupta, S.~Sen and A.A.~Sen, \emph{{GCG Parametrization for Growth Function
  and Current Constraints}},
  \href{https://doi.org/10.1088/1475-7516/2012/04/028}{\emph{JCAP} {\bfseries
  04} (2012) 028} [\href{https://arxiv.org/abs/1110.0956}{{\ttfamily
  1110.0956}}].

\bibitem{Chen:2018dbv}
L.~Chen, Q.-G.~Huang and K.~Wang, \emph{{Distance Priors from Planck Final
  Release}}, \href{https://doi.org/10.1088/1475-7516/2019/02/028}{\emph{JCAP}
  {\bfseries 02} (2019) 028}
  [\href{https://arxiv.org/abs/1808.05724}{{\ttfamily 1808.05724}}].

\end{thebibliography}\endgroup

\end{document}